# Maintaining the Ephemeris of 20 CoRoT Planets: Transit Minimum Times and Potential Transit Timing Variations

Hans J. Deeg[1,2], Peter Klagyivik[1,2,3], J.D. Armstrong[4], David Nespral[1,2], Lev Tal-Or[5], Roi Alonso[1,2], Richelle Cabatic[6], Cameron Chaffey[7], Bartek Gauza[8], Sergio Hoyer[10], Christopher J. Lindsay[11], Paulo Miles-Páez[9], Patricio Rojo[12], Brandon Tingley[13]

[1]Instituto de Astrofísica de Canarias. C. Vía Láctea S/N, E-38205 La Laguna, Tenerife, Spain
[2]Universidad de La Laguna, Dept. de Astrofísica, E-38206 La Laguna, Tenerife, Spain
[3]Institute of Planetary Research, German Aerospace Center, Rutherfordstrasse 2, D-12489 Berlin, Germany
[4]University of Hawai'i Institute for Astronomy, 34 Ohia Ku Street, Pukalani, HI 96768, USA
[5]Department of Physics, Ariel University, Ariel 40700, Israel
[6]Dartmouth College, Thayer School of Engineering, 14 Engineering Drive, Hanover, NH 03755, USA
[7]Department of Physics, University of California, Davis, CA 95616, USA
[8]Centre for Astrophysics Research, School of Physics, Astronomy and Mathematics, University of Hertfordshire, College Lane, Hatfield AL10 9AB, UK
[9]European Southern Observatory, Garching, Germany
[10]Aix Marseille Univ., CNRS, CNES, LAM, Marseille, France
[11]Yale University Department of Astronomy, 52 Hillhouse Ave, New Haven, CT 06511, USA
[12]Department of Astronomy, Universidad de Chile
[13]Bøggildsvej 14, 8530 Hjortshøj, Denmark

## Abstract

We present 33 transit minimum times of 20 transiting planets discovered by the CoRoT mission, which have been obtained from ground-based observations since the mission´s end in 2012, with the objective to maintain the ephemeris of these planets. Twelve of the observed planets are in the CoRoT fields near the galactic center and the remaining eight planets are in the fields near the anticenter. We detect indications for significant transit timing variations in the cases of CoRoT 3b, 11b, 13b, 27b. For two more planets (CoRoT 18b and 20b) we conclude that timing offsets in early follow-up observations led to ephemeris in discovery publications that are inconsistent with timings from follow-up observations in later epochs. In the case of CoRoT-20b, this might be due to the influence from a further non-transiting planet. We also note that a significant majority (23 of 33) of our reported minimum times have negative O-C values, albeit most of them are within the expected uncertainty of the ephemeris.

## 1. Introduction

CoRoT was the first satellite mission with a principal dedication to extrasolar planets (Baglin et al. 2006, Auvergne et al. 2009), having led to the discovery of 37 transiting planets to date (Deleuil et al. 2018, with Moutou et al. 2013 for a more detailed overview over the first 23 planets). The mission was active in the years 2008-2012 and pointed to 24 different fields which were all within two circular zones with a radius of about 7 degrees, called the 'CoRoT Eyes'. One of them was near the galactic center (centered at 18h 50min, 0° in equatorial coordinates) and the other one near the anticenter (at 6h 50min, 0°). The 24 pointings acquired during the lifetime of CoRoT had durations of varying lengths, of 24 to 153 days, and the precision of the ephemeris predicting the times of future transit events is limited accordingly. In particular, planets detected during the short pointings or planets with transits of low signal-to-noise might become 'lost' within a few years, due to uncertainties in the timing of transits that are exceeding 3 hours (Deeg et al. 2015; see also Dragomir et al. 2020 for a similar concerns regarding the current TESS and the previous Kepler/K2 missions). This error was considered the maximum permissible in order to observe a transit reliably during a night with a predicted transit. Given this danger of future transits of the CoRoT planets becoming unobservable in practice, but with the objective to revise the CoRoT planets for the presence of eventual transit timing variations (TTVs), two projects to re-observe their transits from ground were initiated. The results from the first one were recently published by Raetz et al. (2019, further on R+19), covering CoRoT-5b, 8b, 12b, 18b, 20b, and 27b. In this contribution we provide further transit timings of all of them (except CoRoT-5b) and of another 16 CoRoT planets, and indicate potential TTVs. We note that that the TESS mission (Ricker et al. 2015) will provide further transit timings which can then be contrasted against the presented ephemeris. CoRoT anticenter planets were observed in TESS sectors 6 and 7 in winter 2018/19 and will be observed again in sectors 33 and 34 scheduled for winter 2020/21, while TESS pointings to the CoRoT center fields are still to be scheduled and will happen at earliest in spring 2022. In particular, we expect that the transit timings presented here – between the CoRoT and eventual TESS observations – will be useful to check if linear ephemeris describe well the transit times or if changing planet orbital periods fit better to the observations. A joint analysis of the ground-based timings presented here and elsewhere, together with those from CoRoT and TESS is the subject of a forthcoming paper (Klagyivik et al., in prep).

## 2. Observations and analysis

The light curves that have been used for the transit times reported here were acquired with a variety of telescopes, as listed in Table 1. Unless indicated otherwise, CCD imaging in R filters was used, with temporal resolutions that were appropriate to the given target, ranging from 10sec to 3 minutes, and the light curves were obtained using the observers' particular photometry software. The extraction of the transit's mid-time $T_c$ from the light curves was however performed consistently by an experienced member of our team (HJD), employing the following considerations:

Usually, ground-based transit timings are being derived from light curves that include both transit ingress and egress. For example, nearly all light curves which are collected in the Exoplanet Transit Database (ETD, Poddany et al. 2010) are from full transits. However, for about half of the cases reported in this communication, the timings are based on partial transits that include only ingress or egress. This difference arose from a combination of limited observing windows and the uncertainty in the predicted transit times. In the cases of incomplete transits, the moments of first ($T_1$) or last contact ($T_4$) were derived by visual inspection of the light curves, given that these contacts are the features in a transit light curve that can most reliably be recognized. The trustworthiness of each determination of $T_1$ or $T_4$ was evaluated from a comparison of our ground-based light curves against those from CoRoT, considering the following factors:

- The overall noise of the light curve and clearness of recognizing an in- or egress
- The slope of the in- or egress in comparison to the CoRoT light curves.
- The time-difference between the observed and the predicted moment of $T_1$ or $T_4$, considering the expected prediction error.
- In cases were a complete in- or egress was observed, the duration of the in/egress and the amplitude of the transit was also evaluated against CoRoT curves.

Only detections considered as secure are included in this communication. For partial transits, their center-times $T_c$ were then derived as :

$$T_c = T_1 + T_{14}/2 \quad \text{or} \quad T_c = T_4 - T_{14}/2 \quad ,$$

where $T_{14}$ is the duration of an entire transit, for which the values that were reported in the planets' discovery publications were used (see Table 2 for references). If our ground-observations included both $T_1$ and $T_4$, the times $T_c$ were derived from averaging $T_1$ and $T_4$. In cases with well recognizable in- and egress slopes or for full transits, $T_c$ was derived using the bisected chord method, unless noted otherwise in Table 1.

Error-estimates of $T_c$ are based on visual estimations of an acceptable range for $T_1$ or $T_4$ values, combined with the errors of $T_{14}/2$ that have been reported in the literature, or - for the full transits - considering the range of acceptable results from the bisected chord method.
It was attempted to write and use a specific pipeline to recognize the moments of $T_1$ or $T_4$ and to determine their values and errors. Due to the large variety of light curves in terms of S/N, transit-coverage and temporal resolution, this effort did however not provide results of sufficiently consistent reliability. Since most results reported here are based only on in- or egress observations, the use of more sophisticated methods for the determination of $T_c$ for the cases of fully observed transits was then also discarded.

## 3. Results and discussion of individual systems

Table 1 lists the transit times $T_c$ that have been observed in this project, together with their error, $\sigma_{Tc}$ , and the type of transit that was observed: I = ingress ($T_1$) , E = egress ($T_4$), B = both an in- and egress was observed at least partially, F= full transit observed. Furthermore, we indicate cycle-numbes and O-C residuals against the ephemeris that are compiled in Table 2. The next column, $S/N_{O-C}$ , is an indicator for the relevance of an O-C residual, in terms of the number of 'sigmas'. The noise N corresponds to the expected uncertainty of the transit time $T_c$ , based on the period and epoch error of a given ephemeris. N is then obtained by the error-sum of the timing measurement error and the uncertainty of the ephemeris:

$$S/N_{O-C} = (O\text{-}C) / \sqrt{(\sigma_{Tc}^2 + \sigma_{eph}^2)}$$

where

$$\sigma_{eph}^2 = \sigma_E^2 + (E\ \sigma_P)^2 ,$$

with $\sigma_E$ and $\sigma_P$ being the ephemeris' epoch and period errors from Table 2, and E being the cycle number. The next column indicates the telescopes used and the rightmost one provides references to further $T_c$ values that we are aware of. For entries from ETD, in some cases (indicated in Table 1) we list only the number of timings with ETD's quality indicator of DQ ≤ 3, meaning good to excellent curves.

Large values of $S/N_{O-C}$ should only be considered as first indicators for potential TTVs; these are discussed in the notes to the individual systems that follow below. $S/N_{O-C}$ is not a reliable indicator for TTV's because the errors of the ephemeris in the literature did not only depend on quantifiable parameters which are relevant for an ephemeris' precision (transit depth, in/egress duration, photometric noise, length of coverage; Deeg 2015 and Deeg & Tingley 2017), but they were also derived using a variety of different methods. This led to significant inconsistencies among their reported errors, as was pointed out by Deeg & Tingley (2017).

Table 1. Observed transit center times. They are indicated in barycenter corrected universal time.

| CoRoT planet | $T_c$ ($BJD_{UTC}$ - 2400000) | $\sigma_{Tc}$ (d) | Type | Cycle (E) | O-C (d) | $S/N_{O-C}$ | Telescope, reference[1] | Reference to further $T_c$ values |
|---|---|---|---|---|---|---|---|---|
| Center field | | | | | | | | |
| 2b | 56855.51267 | 0.00036 | F | 1502 | -0.0035 | -1.4 | IAC80 | >90 $T_c$ in ETD |
| 3b | 57986.46160[2] | 0.00130 | B | 870 | -0.0938 | -20.6 | GTC | |
| 8b | 56885.53000 | 0.00300 | B | 426 | -0.0047 | -1.1 | IAC80 | 2 $T_c$ in R+19<br>2 $T_c$ in ETD (DQ≤3) |
| 9b | 56889.89534 | 0.00330 | E | 16[3] | 0.0056 | 1.6 | LCO 2M-FTN<br>LCO 1M-SSO | |
| 10b | 56868.43692 | 0.00700 | E | 196 | -0.0643 | -1.6 | IAC80 | |
| 11b | 56828.41800 | 0.00500 | E | 745 | -0.0368 | -3.8 | WISE 0.46m | 3 $T_c$ in ETD |
| 16b | 56834.63000 | 0.00800 | F | 357 | -0.0442 | -0.6 | IAC80 | |
| | 56861.38200 | 0.00800 | F | 362 | -0.0535 | -0.7 | LCO 1M-SAAO | |
| 17b | 56852.49700 | 0.01000 | F | 512 | -0.0795 | -0.5 | LT | |
| 27b | 56810.47029 | 0.00500 | E | 297 | -0.0837 | -4.5 | IAC80 | |
| | 56853.37729 | 0.00500 | E | 309 | -0.0806 | -4.2 | STELLA | |
| 29b | 56075.23000 | 0.00700 | F | 113 | 0.0006 | 0.1 | LCO 2M-FTN | 2 $T_c$ in Pallé et al. (2016) |
| | 56853.43500 | 0.00500 | B | 386 | 0.0000 | 0.0 | IAC80 (Cabrera et al. 2015) | |
| 30b | 56861.48000 | 0.00500 | F | 132 | 0.0388 | 1.2 | IAC80 (Bordé et al. 2020) | |
| 36b | 55814.17090 | 0.00400 | E | 28 | 0.0032 | 0.8 | WISE 1m | |
| | 56864.45890 | 0.00500 | E | 215 | 0.0000 | 0.0 | IAC80 | |
| Anticenter field | | | | | | | | |
| 4b | 57021.48092 | 0.00700 | E | 313 | -0.1249 | -1.1 | IAC80 | |
| 12b | 56997.60750 | 0.00500 | E | 919 | 0.0098 | 0.8 | IAC80 | 3 $T_c$ in R+19<br>6 $T_c$ in ETD |
| | 57099.41510 | 0.00110 | F | 955 | 0.0079 | 0.6 | LT | |
| 13b | 57046.42800 | 0.00300 | F | 559 | -0.0523 | -3.1 | IAC80 | |
| 14b | 57019.54396 | 0.01000 | F | 1476 | -0.0441 | -0.2 | IAC80 | |
| | 57084.57000 | 0.00300 | F | 1519 | -0.0401 | -0.2 | Dan. 1.54m | |
| | 57087.59500 | 0.00400 | F | 1521 | -0.0393 | -0.2 | Dan. 1.54m | |
| 15b | 57061.06708 | 0.01000 | I | 754 | -0.0052 | -0.2 | LCO 1M-SSO | |
| | 57789.42240[2] | 0.00090 | F | 992 | -0.0155 | -0.5 | GTC | |
| 18b | 55589.63389[4] | 0.00000 | F | 141 | -0.0044 | -17.8 | Euler 1.2m (Hebrard et al. 2011) | 4 $T_c$ in R+19<br>3 $T_c$ in ETD (DQ≤3) |
| | 57056.50700 | 0.00300 | F | 946 | -0.0008 | -0.3 | IAC80 | |
| 20b | 55515.55635 | 0.00200 | I | 27 | -0.0111 | -4.9 | WISE 1m (Deleuil+12) | 2 $T_c$ in R+19<br>3 $T_c$ in ETD |
| | 56633.99030 | 0.00200 | F | 148 | -0.0019 | -0.7 | LCO 2M-FTN | |
| 37b | 55913.57646 | 0.01030 | I | 3 | -0.0048 | -0.5 | WISE 0.46m | |
| | 55953.67150 | 0.00390 | I | 5 | 0.0006 | 0.2 | IAC80 | |
| | 56334.52250 | 0.00320 | I | 24 | 0.0000 | 0.0 | IAC80 | |

Notes:

[1] The full names of the telescopes are provided in the acknowledgements. A reference is only given if the observation has been reported previously.

[2] Light curve obtained from white-light fluxes of a time-series of spectra taken with the GTC's OSIRIS instrument using the R1000R grism. $T_c$ and its error was derived from a multi-parameteric fit of the transit (Nespral 2019)

[3] Cycle number in the ephemeris by Bonomo et al. (2017), which is based on a reobservation by Spitzer on 2010 Jun 18. The cycle number would be 24 in the original ephemeris by Deeg al. al (2010), which counts from the first CoRoT transit. See also discussion of CoRoT-9b.

[4] Reconstructed $T_c$ value, based on the ephemeris by Hébrard et al. (2011) and a light curve from the Euler 1.2m provided in the same paper, see text to CoRoT-18b.

Table 2. Ephemeris of planets mentioned in this work.

| Target (CoRoT -NNb) | $T_0$ (BJD) | $T_0$ err (d) | $P$ (d) | $P$ err (d) | Source |
|---|---|---|---|---|---|
| Center field | | | | | |
| 2b | 54237.53562 | 0.00014 | 1.7429964 | 1.7E-06 | Alonso et al. 2008 |
| 3b* | 54283.13940 | 0.00030 | 4.2568000 | 5.0E-06 | Deleuil et al. 2008 |
| 3b | 54283.13388 | 0.00025 | 4.2567994 | 3.5E-06 | Triaud et al. 2009 |
| 8b | 54239.03317 | 0.00049 | 6.2124450 | 7.0E-06 | R+19 |
| 9b | 54603.34470 | 0.00010 | 95.2738000 | 1.4E-03 | Deeg et al. 2010 |
| 9b* | 55365.52723 | 0.00037 | 95.2726560 | 6.8E-05 | Bonomo et al. 2017 |
| 10b* | 54273.34360 | 0.00120 | 13.2406000 | 2.0E-04 | Bonomo et al. 2010 |
| 10b | 54273.34360 | 0.00120 | 13.2402720 | 3.6E-05 | $T_0$: Bonomo et al. 2010<br>$P$: this work |
| 11b | 54597.67900 | 0.00030 | 2.9943300 | 1.1E-05 | Gandolfi et al. 2010 |
| 16b | 54923.91380 | 0.00210 | 5.3522700 | 2.0E-04 | Ollivier et al. 2012 |
| 17b | 54923.30930 | 0.00360 | 3.7681000 | 3.0E-04 | Csizmadia et al. 2011 |
| 27b | 55748.68400 | 0.00100 | 3.5753200 | 6.0E-05 | Parviainen et al. 2014 |
| 29b* | 55753.11500 | 0.00100 | 2.8505700 | 6.0E-06 | Cabrera et al. 2015 |
| 29b | 55753.11500 | 0.00100 | 2.8505616 | 7.2E-06 | $T_0$: Cabrera et al. 2015,<br>$P$: Pallé et al. 2016 |
| 30b | 55665.51460 | 0.00120 | 9.0600500 | 2.4E-04 | Bordé et al. 2020 |
| 36b | 55656.90480 | 0.00049 | 5.6165307 | 2.3E-05 | $T_0$: S. Grziwa (priv.com.)<br>$P$: this work |
| Anticenter Field | | | | | |
| 4b | 54141.36416 | 0.000890 | 9.20205000 | 3.7E-4 | Aigrain et al. 2008 |
| 7b | 54398.07756 | 0.000600 | 0.85359159 | 6.0E-7 | Barros et al. 2014 |
| 12b* | 54398.62707 | 0.000360 | 2.82804200 | 1.3E-5 | Gillon et al. 2010 |
| 12b | 54398.62771 | 0.000240 | 2.82805268 | 6.5E-7 | R+19 |
| 13b | 54790.80910 | 0.000600 | 4.03519000 | 3.0E-5 | Cabrera et al. 2010 |
| 14b | 54787.66940 | 0.005300 | 1.51214000 | 1.3E-4 | Tingley et al. 2011 |
| 15b | 54753.56080 | 0.001100 | 3.06036000 | 3.0E-5 | Bouchy et al. 2011 |
| 18b | 55321.72412 | 0.000180 | 1.90006930 | 2.8E-6 | Hébrard 2011 |
| 18b* | 55321.72565 | 0.000240 | 1.90009000 | 5.0E-7 | R+19 |
| 20b | 55266.00010 | 0.001400 | 9.24285000 | 3.0E-4 | Deleuil et al. 2012 |
| 20b* | 55266.00160 | 0.001000 | 9.24318000 | 9.0E-6 | R+19 |
| 24b | 54789.61100 | 0.006000 | 5.11340000 | 6.0E-4 | Alonso et al. 2014 |
| 24c | 54795.38030 | 0.026500 | 11.75900000 | 6.3E-3 | Alonso et al. 2014 |
| 37b | 55853.44678 | 0.000330 | 20.04482300 | 1.3E-4 | $T_0$: D. Gandolfi (priv. comm.), *P*: this work |

* If more than one ephemeris is given, the starred one is used for the O-C residuals of Table 1.

Below we provide comments to all systems observed, in the order of their listing in Table 1 . Plots of light curves for all timing measurements of Table 1 (except for previously published curve by the Euler 1.2m of CoRoT-18b, which was not obtained by our team) are shown in the Appendices, whereas the corresponding tabulated light curves will become available at the VizieR service of the Strasbourg astronomical Data Center (CDS).

CoRoT-2b: For this system, ETD lists currently timings from over 90 follow-up observations, with more than half of them being considered of good to excellent data quality, using ETD's data quality (DQ) indicator of 3 or lower as reference. All of these timings line up very well and within on O-C of ±0.01 d with the ephemeris of Alonso et al. (2008), of which our measurement is no exception. CoRoT-2b counts also with a few which pre-discovery timings obtained about 2 years before the CoRoT observations (Rauer et al. 2010). Our light curve taken with the IAC80 (Fig. 1) has been analyzed many times in a university course using the TAP (Gazak et al. 2012) transit analyzer with a multi-parametric MCMC chain, from which the $T_c$ value in Table 1 has been derived.

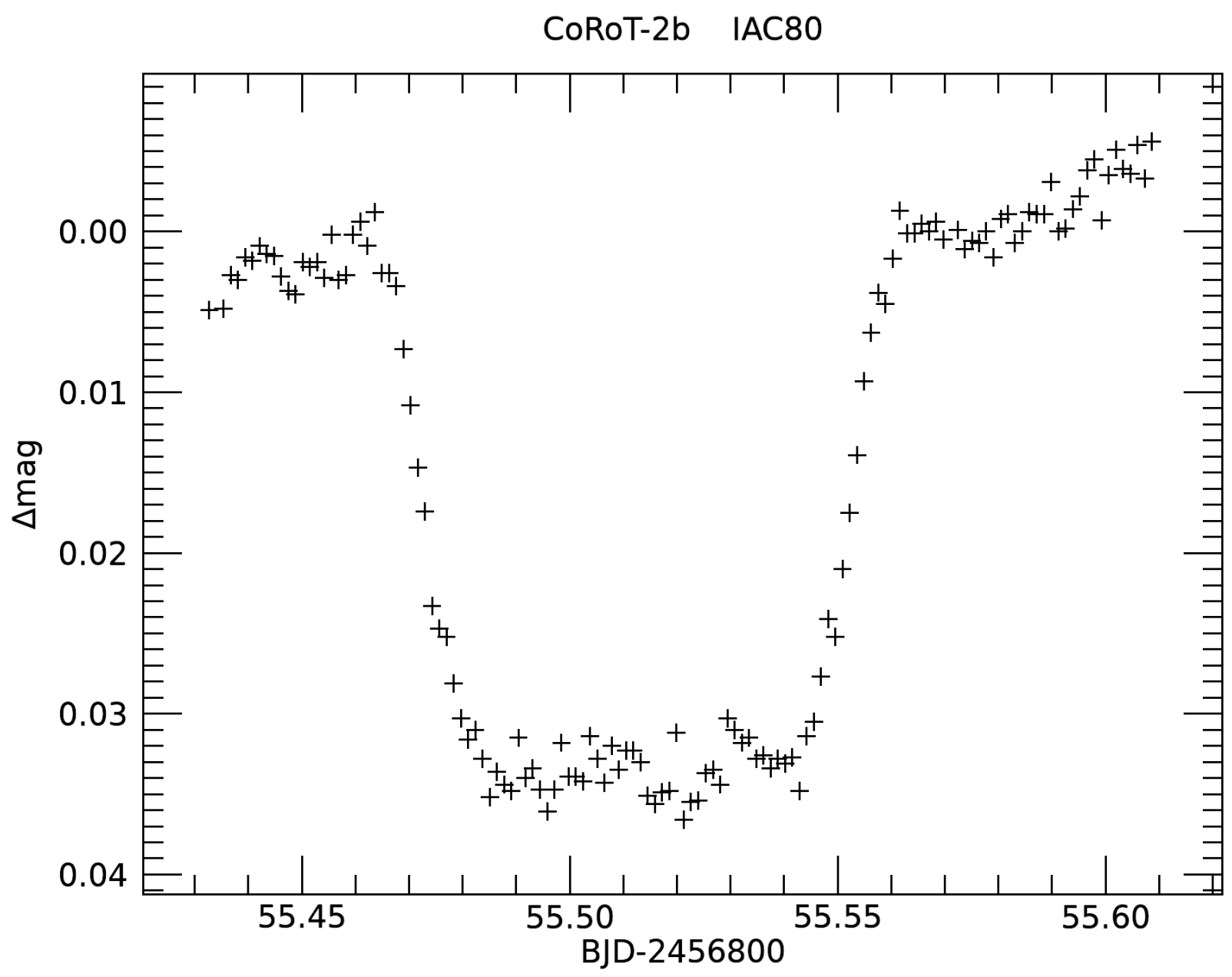

Fig. 1. Light curve of a transit of CoRoT-2b acquired with the IAC80 on 2014 Jul 16. The vertical axes is in uncalibrated relative magnitudes. This plot is similar to those shown in the Appendices for all transits observed for this work.

CoroT-3b: In data acquired with the 10.4m Gran Telescopio Canarias (GTC) on 2017 Aug 20, the transit appears 2.25 hours earlier than predicted from the ephemeris of Deleuil et al. (2008). An alternative ephemeris derived from the same original CoRoT data by Triaud et al. (2009) leads to a similar offset, with the GTC transit being 2.11h too early. In either case, the deviation in the transit time is much larger than the uncertainty of Corot-3b's ephemeris, which was ±6 resp.± 4 minutes in Aug 2017. The ETD database provides three timing values taken in 2009, 2010 and 2017, which do not indicate any deviation in periodicity. A revision of the underlying light curves in ETD led us however to the conclusion that these are of too low quality for the provision of meaningful estimates of the transit times, as they lack any well recognizable partial or full transits. This target does therefore exhibit likely transit timing variations and should be re-observed with priority, with results from TESS being awaited.

Corot-8b: A first analysis of our transit times showed a significant deviation against the ephemeris published in CoRoT-8b's discovery paper (Bordé et al. 2010), which incidentally claimed the potential presence of transit timing variations. An error in the ephemeris by Bordé et al. was then found, with its $T_0$ being ~85 minutes earlier than the first transit in the CoRoT data. Also, R+19 published a revised ephemeris based on their own follow-up observations (Table 1) plus the full set of CoRoT transits. Against these revised ephemeris, our transit-timing acquired with the IAC80 is within the expected uncertainties.

CoRoT-9b: After discovery of this planet (Deeg et al. 2010), a further transit was observed by CoRoT itself in a dedicated pointing on 2011 Jul 4, which was observed simultaneously by the Spitzer mission in the 4,5µm band (Bonomo et al. 2017). The mid-transit times $T_c$ differ between the CoRoT and the Spitzer observation by only 104 seconds, which implies that transit mid-times have little dependence on the wavelength. These observations covered a baseline between the first and last transit of 3.1 years and permitted Bonomo et al. the derivation of an ephemeris of improved precision. This ephemeris has however an epoch ($T_0$) that was reset to another transit that they observed with Spitzer, on 2010 Jun 18.

Due to the long orbital period, the transits of CoRoT-9b last 8.1 hours with in/egresses of about 1 hour, implying that transit features are difficult to detect due to the slowly varying flux-levels. The light curve of a transit on 2014 Aug 20 was acquired first with the 2m LCO telescope on Mt. Haleakala, Hawaii, followed by the 1m LCO telescope at Siding Springs, Australia, using in both cases a PanSTARRS i-band filter. While the 2m telescope generated a featureless flat light curve - having fallen completely into the central part of the transit - the curve from the 1m telescope showed an egress, which was modelled in detail, using the UFIT/UTM transit modeler (Deeg 2014). This software employs a Monte-Carlo Markov Chain (MCMC) algorithm, in which we kept the transit model fixed to the values given in the CoRoT-9b discovery paper (Deeg et al. 2010) while leaving only three parameters free. These were the mid-time of the transit and the offset and slope of the off-transit flux-level as a function of time. The best shows an excellent agreement between the model and the data (Fig. 2), and indicates a $T_c$ that is only 8 minutes later than predicted by the ephemeris of Bonomo et al. (2017).

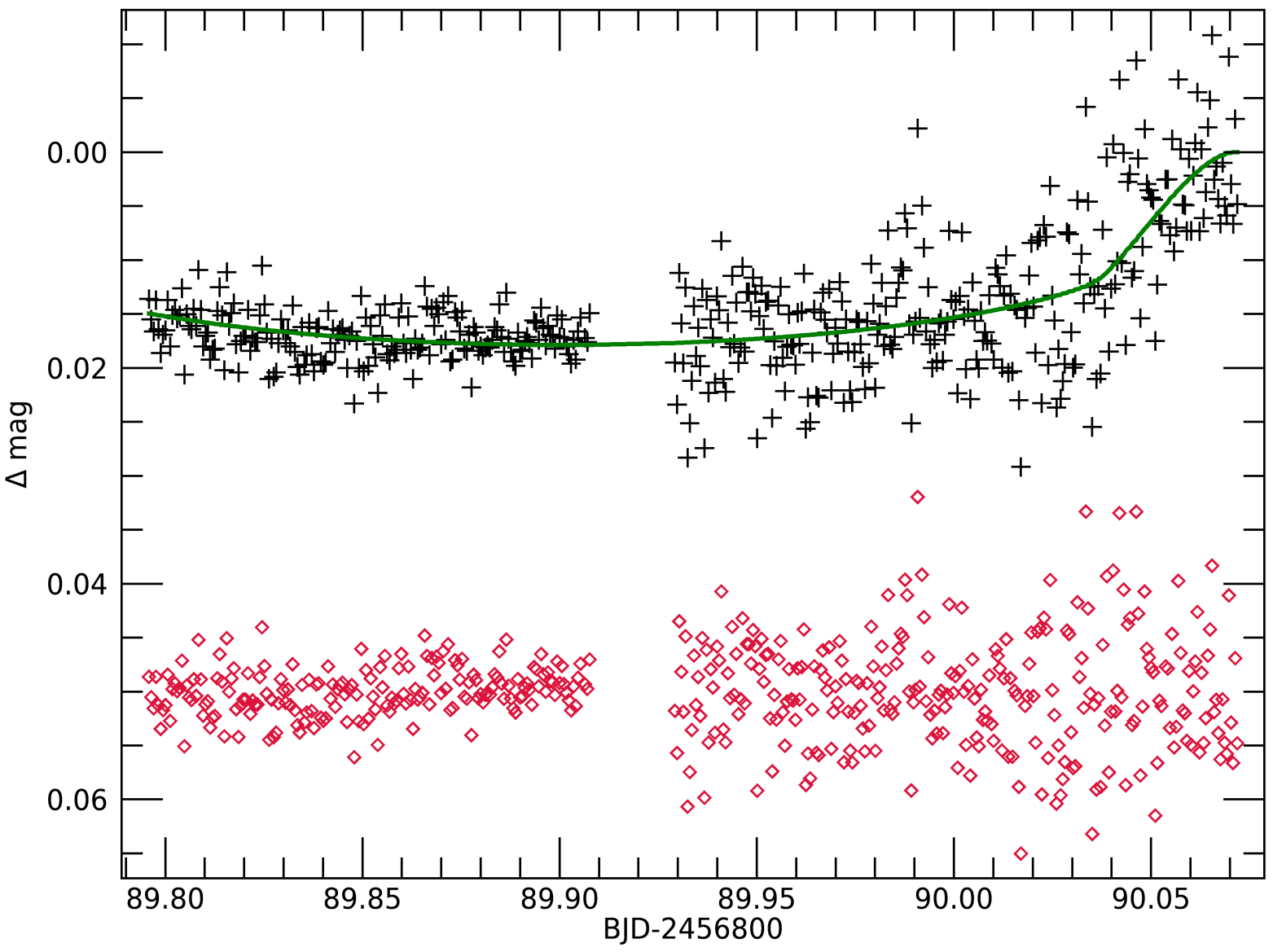

Fig. 2. Light curve of a partial transit of CoRoT-9b observed on 2014 Aug 20 with the 2m LCO telescope at Haleakala Observatory, Hawaii and the 1m LCO telescope at Siding

Springs, Australia (black crosses; left section from the 2m and right one from the 1m), with the best fit of a transit-model (green line) to the 1m data. A slight slope in the LCO 1m data that was originally present has been removed from both data and model fit. The 2m LCO observations were not used in the fit but were shifted in $\Delta$ mag for an optimal agreement with the transit model. The red symbols are the residuals, which are offset downwards by 0.05mag for better visibility.

CoRoT-10b: Our $T_c$ value listed in Table 1 is the first successful reobservation of CoRoT-10b (discounting an unreliable entry in ETD) and shows a moderate 1.6 sigma deviation from the original ephemeris by Bonomo et al. (2010). 10b was one of the CoRoT planets which was in the danger of getting 'lost' (Deeg et al. 2015) and our re-observation permits a dramatic increase in the precision of its ephemeris. A new derivation of the period has therefore been included in Table 2.

CoRoT-11b: Relative to the ephemeris in the discovery paper by Gandolfi et al. (2010), our timing from 2014 June 19 taken with the 0.46m telescope of WISE observatory, Israel, is early by 53 minutes, which is 3.8 times larger than the uncertainty implied by that ephemeris. The ETD and TRESCA databases contain three further transits of good quality taken in 2011 and 2012, which corroborate transit times that are about 4 sigma earlier than implied by the Gandolfi et al. ephemeris. A revision of that ephemeris, which was entirely based on CoRoT transits acquired between 2008 Apr. 15 and Sep. 7, does not reveal any source for this discrepancy. CoRoT-11b might therefore be a case of a real transit timing variation.

CoRoT-16b: The light curves underlying our two timing measurements in Table 1, if taken individually, would not have been of sufficient quality for inclusion into that table, given their noisiness which is due to the target's faintness (Rmag=15.5). An overlay of both light curves (Fig. 3) indicates however a good agreement between each other, showing a correct transit duration of 0.1d and depth of 1%, therefore warranting their inclusion. We note that the transits occur about 0.05d or 70min before the predicted transit times, using the ephemeris of the CoRoT-16b discovery paper (Ollivier et al. 2012). This deviation is however smaller than the ephemeris' 1-sigma uncertainty of ±103 minutes at the epoch of our observations.

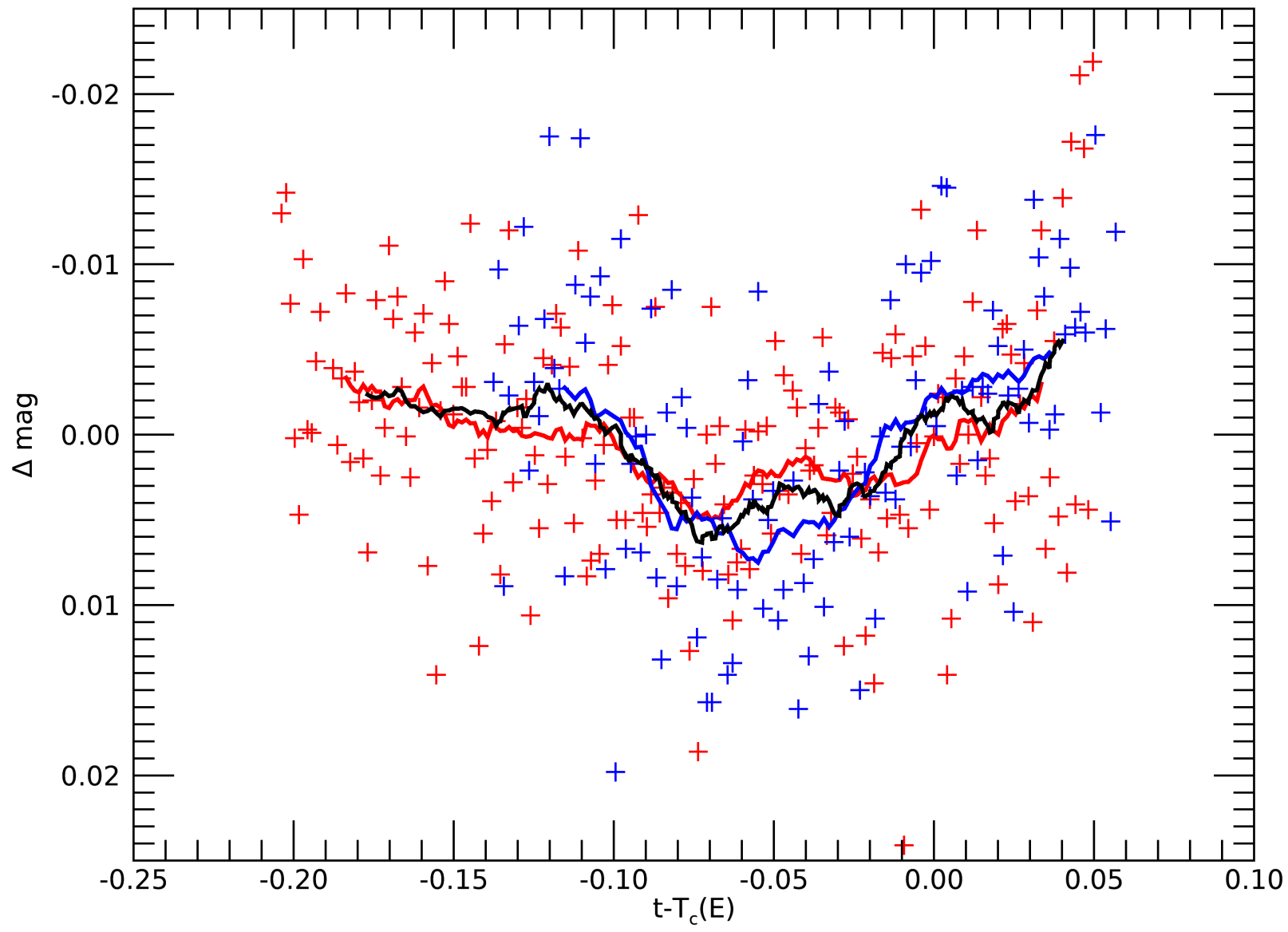

Fig. 3. Superposition of light curves of CoRoT-16b transits observed on 2014 Jun 26 (red crosses ) with the IAC80 and on 2014 Jul 22 (blue crosses) with the LCO 1m at SAAO, in this case using a PanSTARRS i-band filter. The red and blue solid lines are boxcar smoothings over 25 points of the individual light curves, while the black line is a smoothing of the combination of both curves. The horizontal axis indicates the time in days, relative to the predicted transit time $T_c$(E) from the ephemeris of Ollivier et al. (2012), with E=357 and 362.

CoRoT-17b: The transits of this planet are difficult to observe, due to their shallowness of ~0.4% and the target's faintness (Rmag=15.3). CoRoT-17b was observed 3 times within a few weeks at the Liverpool 2m telescope, on 2014 Jun 9, Jul 13 and Jul 28. Similar to CoRoT-16b, the individual transits where not of sufficient quality, but a combination of them (Fig. 4) shows a feature that is reasonably close to the expected transit duration (4.7h = 0.195d) and depth (0.4%) to be considered a very likely detection. The value of $T_c$ given in Table 1 was derived from the combined light curve (black line in Fig. 3) and was assigned to 2014 Jul 13, which was the best of the three data sets. This $T_c$ is 1.9h earlier than indicated by the ephemeris from CoRoT data (Csizmadia et al. 2011), but is well within the ephemeris' uncertainty of ±3.7h at the epoch of our observations.

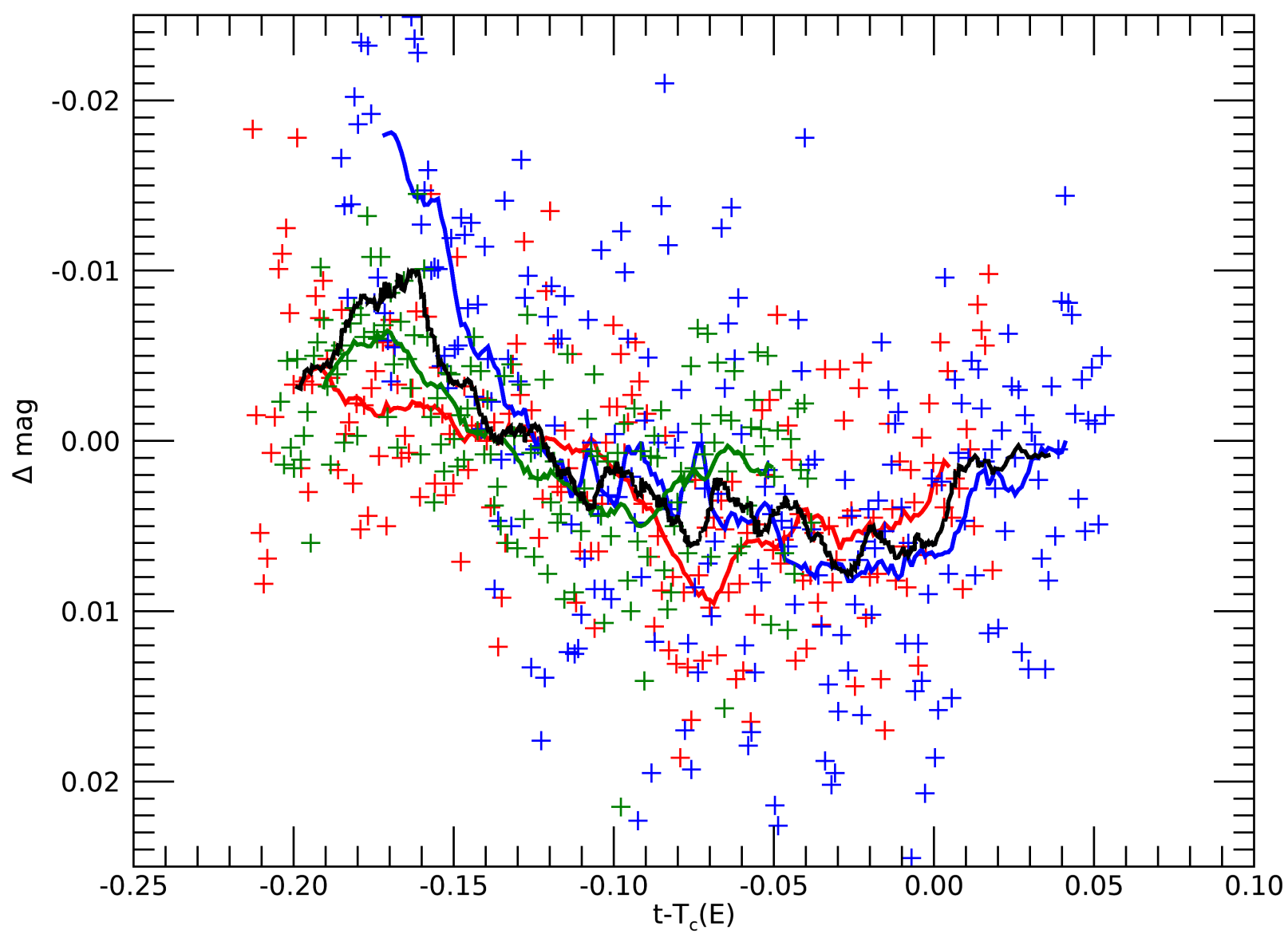

Fig. 4. Similar to Fig.3, showing transits of CoRoT-17b observed on 2014 Jun 9 (red), 2014 Jul 13 (blue) and 2014 Jul 28 (green), with the combined smoothed light curve in black. In all three nights, the airmass was increasing during the observation, which is the likely source for the general slope that is common to all three data sets.

CoRoT-27: Both our light curves (Fig. 5), acquired on 2014 Jun 1 and 2014 Jul 14, show a likely detection of an egress that is about 2.0h earlier than predicted by the discovery paper's ephemeris (Parviainen et al. 2014), corresponding to a 4.5 sigma deviation against the ephemeris' uncertainty of ±25 min at that epoch. R+19 report two later observing attempts from June 2016, which did not detect the transit at all. From this non-detection they conclude that 'the transit must have happened at least 3.9 h earlier or 4.5 h later' (relative to Parviainen's ephemeris). If we extrapolate our deviation of 2.0h to the epoch of R+19's observations, they should have detected the transits at 3.3h earlier, well within their observing window. We therefore expect that CoRoT 27b has a notable transit timing variation with an increasingly non-linear offset relative to Parviainen's ephemeris.

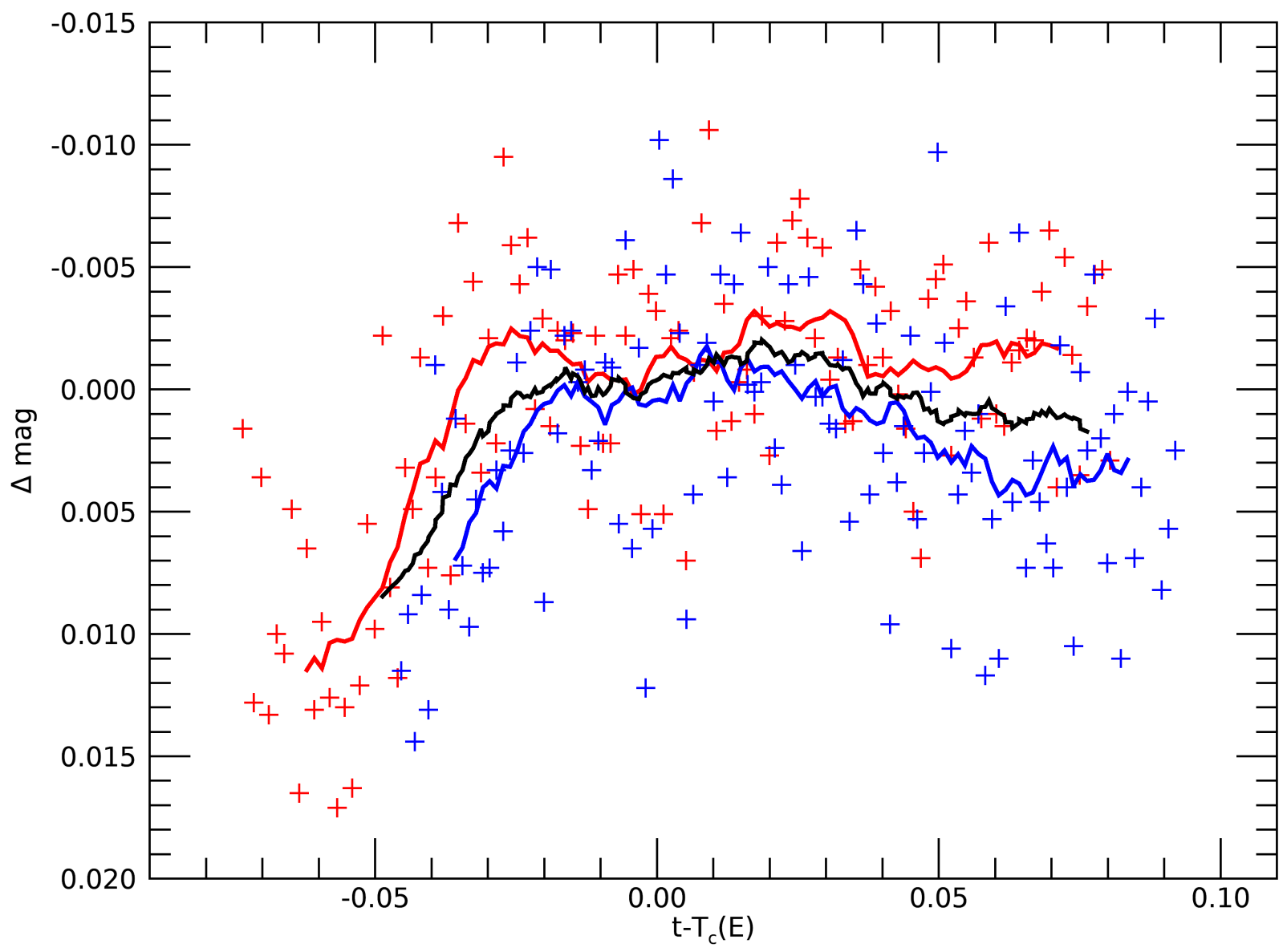

Fig. 5. Similar to Fig.3, but for CoRoT-27b observed with the IAC80 on 2014 June 1 (red) and the STELLA 1m telescope on 2014 July 14 (blue), with the combined smoothed light curve in black. Both transits are significantly earlier than predicted from the ephemeris of Parviainen et al. (2014), causing coverage of the egress only. The smoothing of the curves has been over 15 points.

CoRoT-29b: This planet was among the targets to be observed for the project reported here, but its follow-up concluded in time for inclusion into the CoRoT-29b discovery publication (Cabrera et al. 2015). The $T_c$ from the IAC80 observations on 2014 Jul14 (E=386) was therefore used in the derivation of the ephemeris by Cabrera et al. A light curve of the observation from the LCO's 2m FTN telescope at E=113 is also shown in that paper, but without quoting any $T_c$, which has therefore been included in Table 1. Two further transit timings, acquired with the GTC on 2014 Jul 31 and 2015 Aug 7 for a spectrophotometric study, have been published by Pallé et al. 2016. From these, they provide an updated orbital period (included in Table 2), which is also in good agreement with our $T_c$ measures.

CoRoT-30b: Transit observations of this planet were acquired within the project reported here, but similar to CoRoT-29b, they arrived in time to have been reported in the planet's recent discovery publication (Bordé et al. 2020). However, Bordé et al.'s principal ephemeris (their Table 6) is only based on model fits to CoRoT data and does not take the IAC80 observation from 2014 Jul 22(E=132) into account. They note however that the inclusion of the $T_c$ from that observation increases the precision of the planet's period, arriving at 9.060347(39) days.

CoRoT-36b: (CoRoT-ID 652345526, UCAC2 34324554 , at 18:31:00.26 +07:11:00.3 J2000) is a Jupiter-sized planet with a period of 5.6 days that has been included among the 37 CoRoT planets that are quoted in the overview paper by Deleuil et al. (2018), although a detailed publication is still pending (Grziwa et al. in prep). The ephemeris given in Table 2 has been determined from a $T_0$ based on CoRoT data (Grziwa, priv. com) and from the IAC80 timing on 2014 Jul 25 (E=215).

CoRoT-4b: Our $T_c$ value obtained from an egress is the first published reobservation of CoRoT-4b (Aigrain et al. 2008, with a more detailed description in Moutou et al. 2008) and is within the expected timing error from the original ephemeris.

CoRoT-12b: For this planet exist numerous ground-based follow-up observations, as its 1.9% deep transits are relatively easy to observe. Considering our $T_c$, those from R+19, and the good-quality ones from ETD (DQ of 3 or better), they are all well described by original ephemeris of Gillon et al. (2010) or by the revised one of R+19. We note that Gillon et al. hinted at potential TTVs with an amplitude of ~1 minute and a period of ~68 days. Unfortunately, the precision of the ground-based follow up is not sufficient to corroborate the further presence of this feature.

CoRoT-13b: The $T_c$ value obtained from the light curve (Fig. 5) is 76 minutes early versus the ephemeris of the discovery paper (Cabrera et al. 2010), which corresponds to 3.1 times its uncertainty at the observation's epoch, indicating potential TTVs.

CoRoT-14b: Three transits of good quality were observed with the IAC80 and the Danish 1.54 m telescope (Fig. 6). They were about 1h earlier than predicted by the discovery ephemeris of Tingley et al. (2011), but are well within the ephemeris' uncertainty of 4.6 to 4.7h at the observations' epochs.

CoRoT-15b: The light curve of an ingress was acquired on 2015 Feb 7 with the 1m LCO telescope at SSO and a nearly complete transit was acquired on 2017 Feb 4 with the 10.4m GTC. The GTC light curve was derived from the white-light summation of spectra that were taken with the R1000R filter for a study of transit spectroscopy (Nespral 2019). Both transits (Fig. 7) agree well with the ephemeris of Bochy et al. (2011).

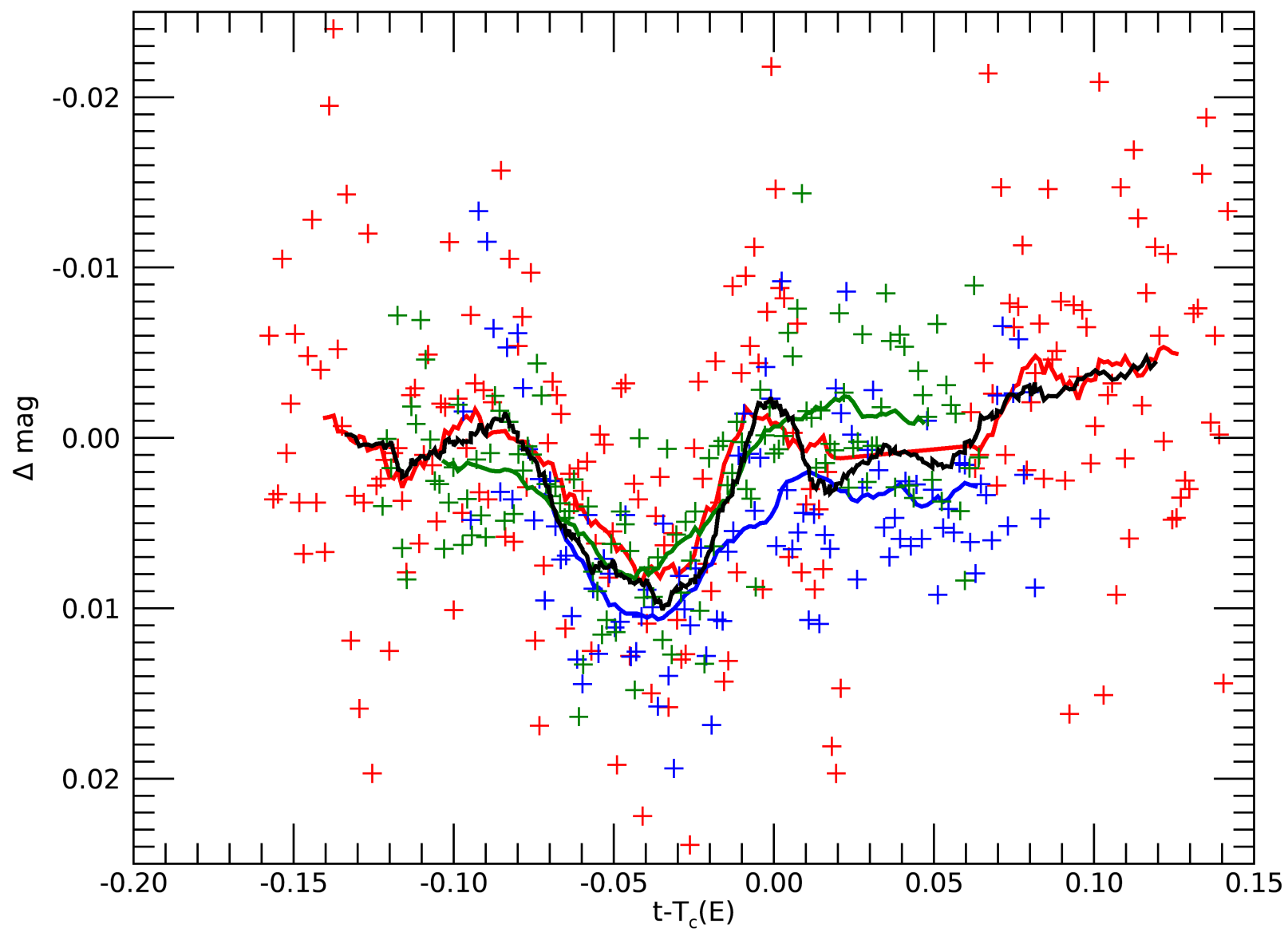

Fig. 6. Similar to Fig.3, showing transits of CoRoT-14b observed with the IAC80 on 2014 Dec 27 (red) and the Danish 1.54m telescope on 2015 Mar 2 (blue) and on 2015 Mar 5 (green), with the combined smoothed light curve in black.

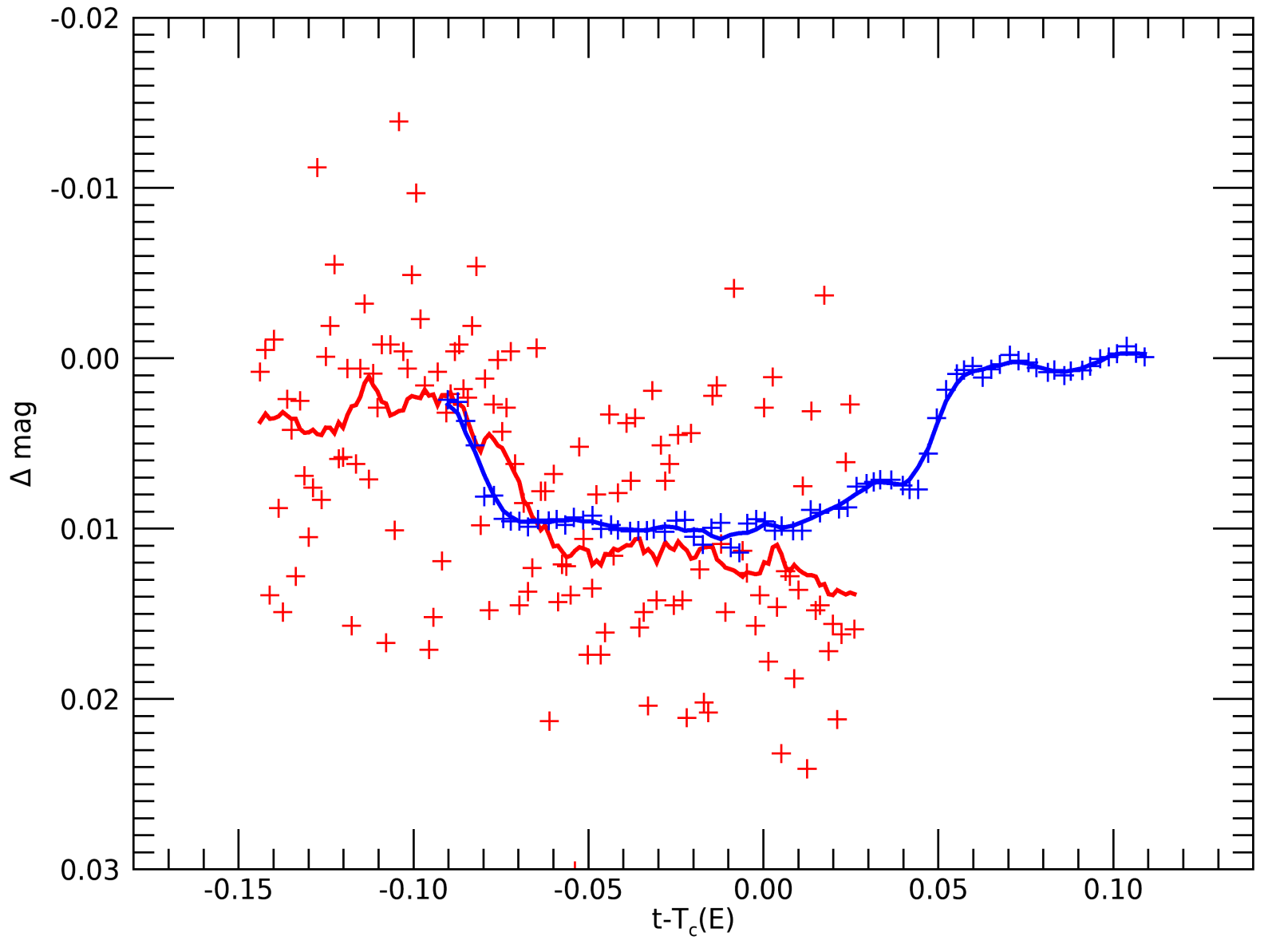

Fig. 7. Similar to Fig.3, showing transits of CoRoT-15b acquired with the 1m LCO telescope (red) and the 10.4m GTC (blue). For the GTC data, the solid line was generated with a boxcar smoothing over only 5 points. Due to the very different noise-characteristics, we refrain from showing the combined curve.

CoRoT-18b: Our transit observed on 2015 Feb 2 (E=946) with the IAC80 is 28 minutes behind the ephemeris in the discovery paper (Hébrard et al. 2011). The very small uncertainty in their quoted period, given the short CoRoT pointing from 2010 Mar 5 to 29, is explained by them from follow-up observations made with the Euler 1.2m telescope about 8 months later (on 2011 Jan 28 at E=141), which were used to refine their ephemeris. We note that for unspecified reasons, Hébrard's ephemeris has a zero-epoch on 2010 May 5, well past the coverage by CoRoT, while the first transit observed by CoRoT corresponds to E=-32.

With the small period-uncertainty by Hebrard et al., the lateness of our IAC80 timing of 28 minutes translates into an error of 7.6 sigma against their ephemeris. However, good-quality entries in ETD (of DQ ≤ 3) as well as the four timings acquired by R+19 all show a similar trend of being late by 7 to 8 sigma against Hebrard's ephemeris. These offsets, both in terms of their absolute sizes and in terms of their significance diminish however greatly if the revised ephemeris of R+19 is employed, against which our IAC80 timing is early by only 1 minute. We note that Hebrard et al. do not indicate the $T_c$ of their Euler observations at E=141, but using their ephemeris (see also their Fig. 3) we can reconstruct its value (see entry in Table1). This $T_c$ is now 6 minutes or 17.8 sigma early against the ephemeris by R+19. However, given that all further published timings, over the range of E=714 to 1865 , agree well with the R+19 ephemeris, the Euler 1.2m timing seems to be an outlier and the presence of significant timing variations is unlikely.

CoRoT-20b: Our timing obtained with the LCO 2m telescope is within 3 minutes of the refined ephemeris of R+19, who had two later timing measurements at their disposal. Three further timings from low-quality light curves are also available in ETD. We note that the original ephemeris of Deleuil et al. (2012) was already based on a ground-based timing taken with the 1m WISE telescope, whose $T_c$ value has been included in Table 1. However, this timing has a significant offset against the ephemeris by R+19. Rey et al. (2018) provide evidence from radial velocity follow-up for a further non-transiting planet *c* with an orbital period of 1675 days on an eccentric orbit. They also imply that planet *c* should induce TTVs on planet *b* whose amplitude of a few minutes would vary with the period of planet *c* (see their Fig. 7). Such variations are at the limit of the precision of the current ground-timings, albeit the poor fit of the WISE timing (with an offset by 16 minutes) against the later timings might be related to planet *c*. A more thorough analysis of all available timings together with those from TESS should be the subject of further work.

CoRoT-37b (CoRoT-ID 617963863, TYC 4792-1886-1) is a planet transiting an F4 star in the young cluster NGC 2232, with an orbital period of 20 days (Gandolfi et al., in prep). It was announced as CoRoT-32b in several conferences in 2013[1] and 2014[2]. Under that denominator it was also mentioned in refereed papers by Guenther et al. (2013) and Hatzes (2014), while a dedicated publication is still pending. In the overview of CoRoT detections by Deleuil et al.

[1] 11th CoRoT week, 2013, http://research.iac.es/congreso/cw11/pages/meeting/view-abstractcd04.html?aid=21
[2] The Space photometry Revolution, CoRoT Symp. 3 – KASC-7, 2014, https://corot3-kasc7.sciencesconf.org/37184

(2018) it is however mentioned as CoRoT-37b. The reason for the change in numbering was a publication by Boufleur et al. (2018), which assigned the name CoRoT-32 to an unrelated system (CoRoT 223977153, UCAC2 34993171). The ephemeris given in Table 2 has been derived from a linear fit using a $T_0$ derived from CoRoT data (Gandolfi, priv. com) and from the $T_c$ of the follow-up observations given in Table 1.

In the following, we comment on a several more CoRoT planets that are not included in Table 1:

Corot-7b: Ground observations of the very shallow (0.032% deep, Legér et al. 2009) transits of this Super-Earth are extremely challenging. They were intended on 2010 Jan 15 and 2013 Jan 15, both times with the 10.4m GTC and the OSIRIS imager, using a strongly defocused point-spread function, without obtaining reliable transit detections. After the initial discovery in mission data acquired between Oct. 2007 and March 2008, CoRoT observed this planet in a further pointing from January to March 2012. An ephemeris from this reobservation was published by Barros et al. (2014), with a greatly improved precision over the original one by Legér et al. (2009).

CoRoT-24b and c: This multiplanet system was never attempted to be re-observed by us, given the unlikely recovery of reliable transits due to their shallowness, of 0.15% for *b* and 0.26% for *c* (Alonso et al., 2014), and the very large timing uncertainties, which in 2014 were already ±5.5 h and ±24h for the two planets.

CoRoT-19b, 22b, 23b, 26b, 31b: Transits of these planets were also observed, but the resulting light curves remained inconclusive, mostly due to being too noisy for the expected transit depth; from showing features that are incompatible with a transit; or from being too short to be of discriminatory value. The remaining CoRoT-planets had failed observations due weather or technical issues or we were unable to schedule their observation.

**Conclusions**

Table 1 provides 33 ground-based timing measurements from 20 exoplanet systems. Of them, six systems have timings with $S/N_{O-C} > 3$, that is, the observed deviation from the ephemeris was more than 3 times the expected uncertainty. We consider four of these systems (CoRoT-3b, 11b, 27b, 13b) to display indications for potential TTVs. For these systems, further timing measurements over longer epochs will be needed to corroborate such a diagnostic. In the other two cases, of CoRoT 18b and 20b, the planets' original ephemeris (Hébrard et al. 2011 and Deleuil et al. 2012, respectively) were based not only on the CoRoT data but also on early ground follow-up timings that are included in Table 1. In both cases, our follow-up at later epochs (and for 18b, also further timings from ETD) provide timings that are consistent with linear ephemeris which R+19 had derived from their own follow-up timings. In these revised ephemeris, the notable outlier is then the early ground-observation that had influenced the discovery ephemeris. In the case of CoRoT-20b, this discrepancy might have arisen from TTVs with amplitudes that vary on time-scales of years and which are induced by a long-periodic non-transiting planet. A more thorough analysis of these case is however required in order to ascertain that the early timing outliers could have been caused by the presence of further planets.

Of further note is that a large majority, 23 out of the 33 entries in Table 1, has timings that are earlier than expected, with negative O-C values. This would correspond to periods that are (or are becoming) shorter than the ephemeris' periods. However, no corresponding systematics in timings from Kepler planets without identified TTVs (see Rowe & Thompson 2015, Holczer et al. 2016, Kane et al. 2019 for planets identified with TTVs) have been reported, while such a trend, if real, should have been found in the Kepler mission data, given Kepler's much longer temporal coverage and higher photometric precision. We surmise therefore that our mostly negative O-C values could be the result of some unrecognized systematics that affected many of the original ephemeris derivations from the CoRoT data.

In all cases, we are awaiting a recovery of transits of most of the CoRoT planets in data from TESS and from future ground and space missions, which will maintain the legacy of the planets that were discovered by the first space mission dedicated to exoplanets.

**Acknowledgements**

This work is based on observations of the following telescopes (provided are also the acronyms used in Table 1): IAC80: 80cm telescope of the Instituto de Astrofísica de Canarias at Teide Observatory, Tenerife, Spain. We thank the night operators at Teide Observatory for the acquisition of several of the listed observations. LCO: 1m telescopes of the Las Cumbres Observatory, at Siding Springs Observatory, Australia (SSO) and at the South African Astronomical Observatory (SAAO) and 2m Faulkes Telescope North (FTN) at Haleakala Observatory, Hawaii . WISE: Tel-Aviv University 0.46m and 1m telescopes at WISE Observatory, Israel. LT: 2m Liverpool Telescope of the Liverpool John Moores University, at the Roque Muchachos Observatory, La Palma, Spain. Dan.1,54m: Danish 1.54m telescope at ESO's La Silla Observatory. GTC: 10.4m GTC telescope at Roque Muchachos Observatory, La Palma, Spain. We thank the staff of GTC for their support during queue observations. STELLA: 1.2 m STELLA-WiFSIP telescope of the Leibniz-Institut für Astrophysik Potsdam, at the Teide Observatory, Tenerife, Spain. Both Teide Observatory and Roque Muchachos Observatory are observatories of the Instituto de Astrofísica de Canarias.
The CoRoT space mission, launched on December 27th 2006, was developed and operated by CNES, with the contribution of Austria, Belgium, Brazil, ESA, Germany, and Spain. We acknowledge the use of the IAS CoRoT Public Archive (http://idoc-corot.ias.u-psud.fr/ ) and the COROT Archive at CAB (https://sdc.cab.inta-csic.es/corotfa ). We acknowledge the use and the usefulness of the ETD (http://var2.astro.cz/ETD ) and TRESCA (http://var2.astro.cz/EN/tresca ) databases of the Czech Astronomical Society. We thank Ankit Rohatgi to make freely available to the community the superb WebPlotDigitizer tool (https://automeris.io/WebPlotDigitizer/), which permitted an efficient recovery of several older time-series that had only been preserved in graphical form.

HD, PK, DN acknowledge support by grants ESP2015-65712-C5-4-R and ESP2017-87676-C5-4-R of the Spanish Secretary of State for R&D&i (MINECO). SH acknowledges CNES funding through the grant 837319.

Appendix A
Transit light curves of CoRoT planets in the galactic center field. They are ordered first by planet number and then by the BJD.

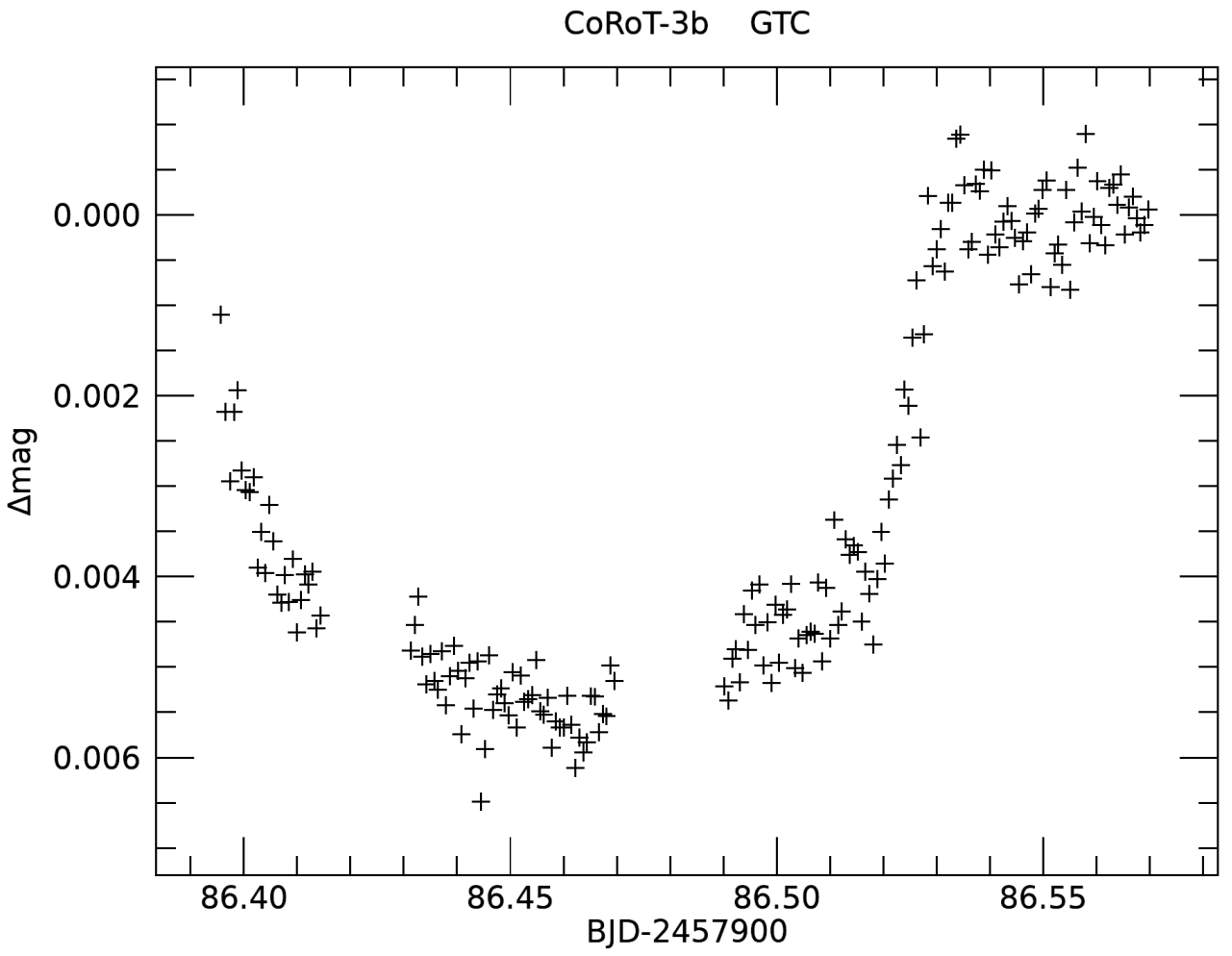

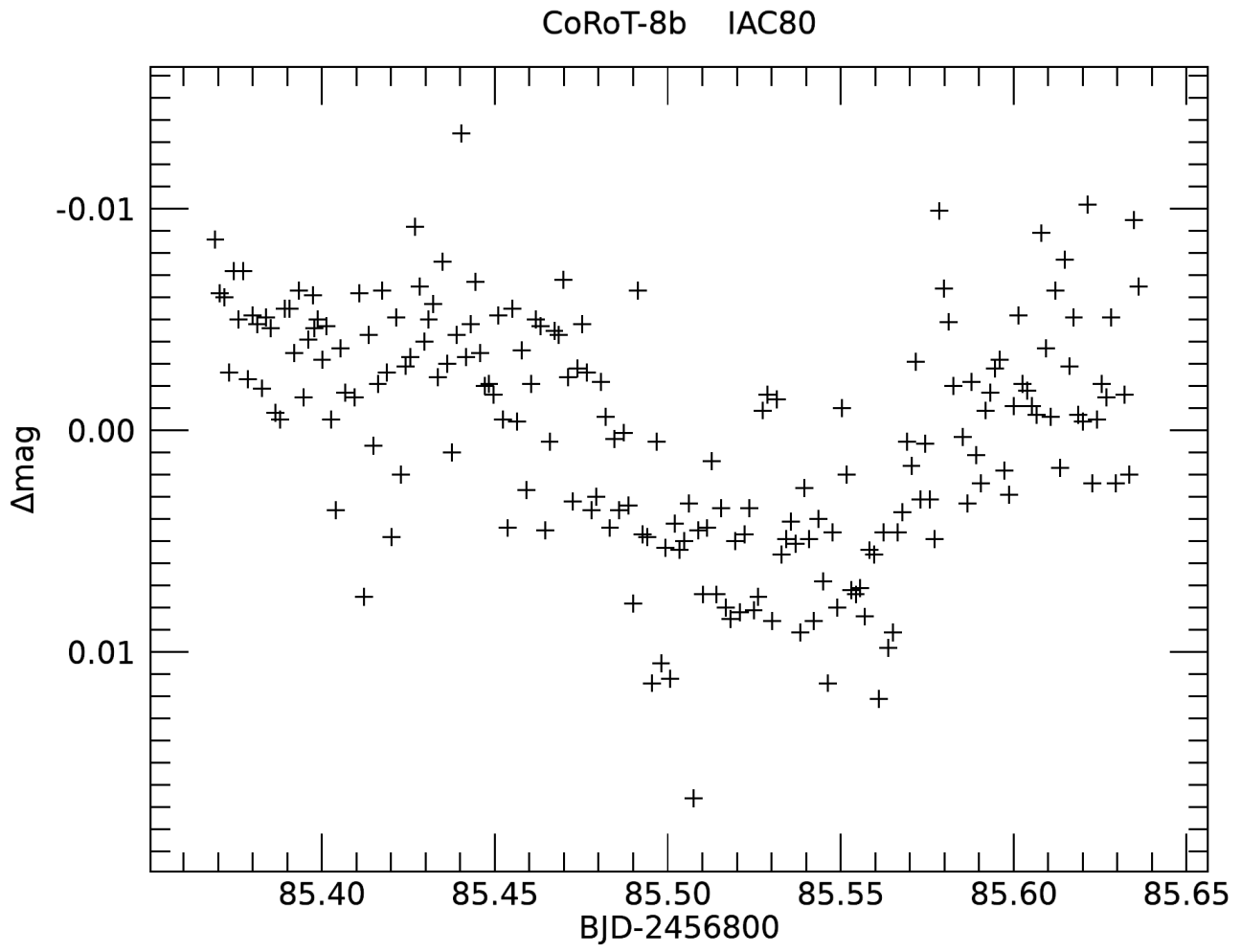

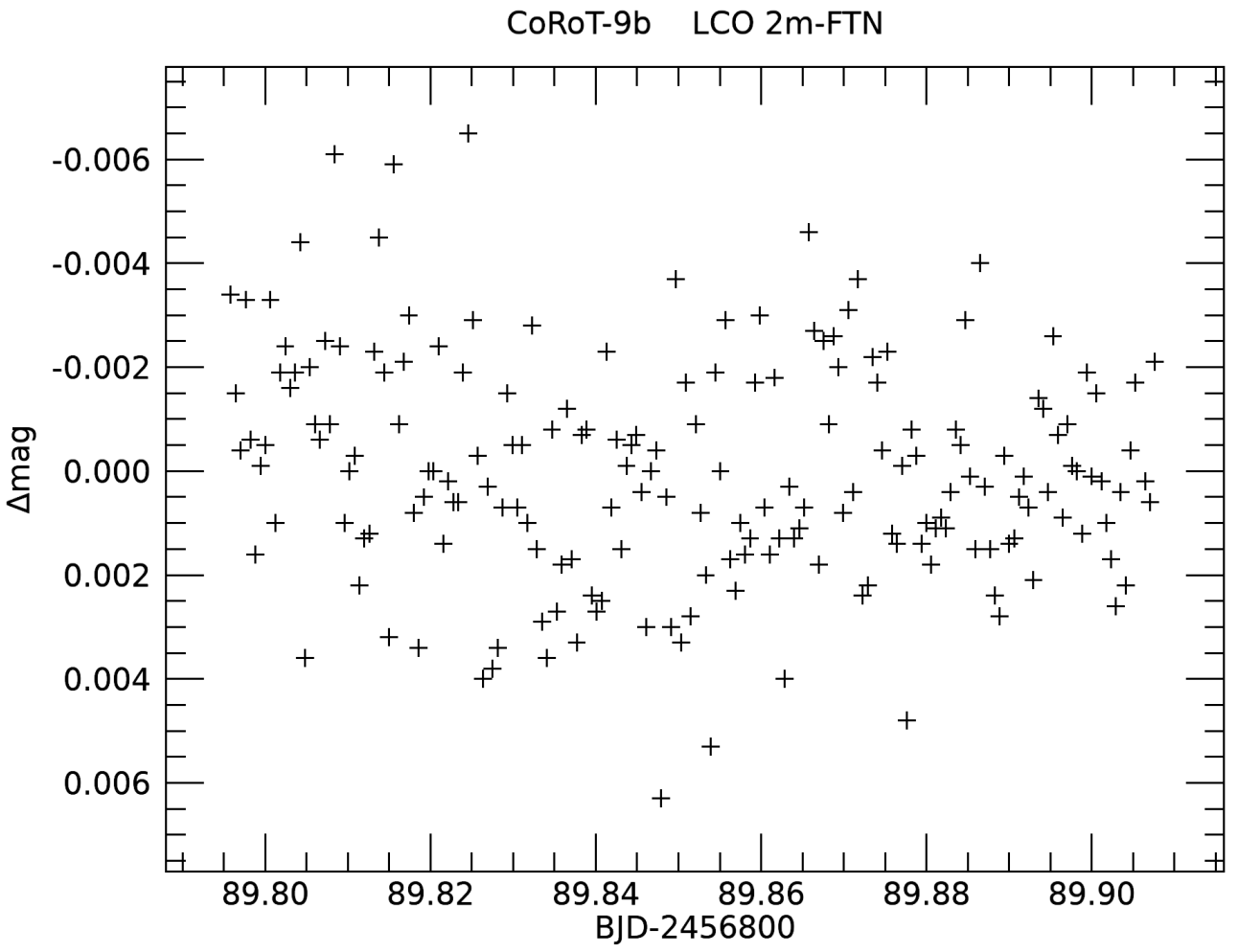

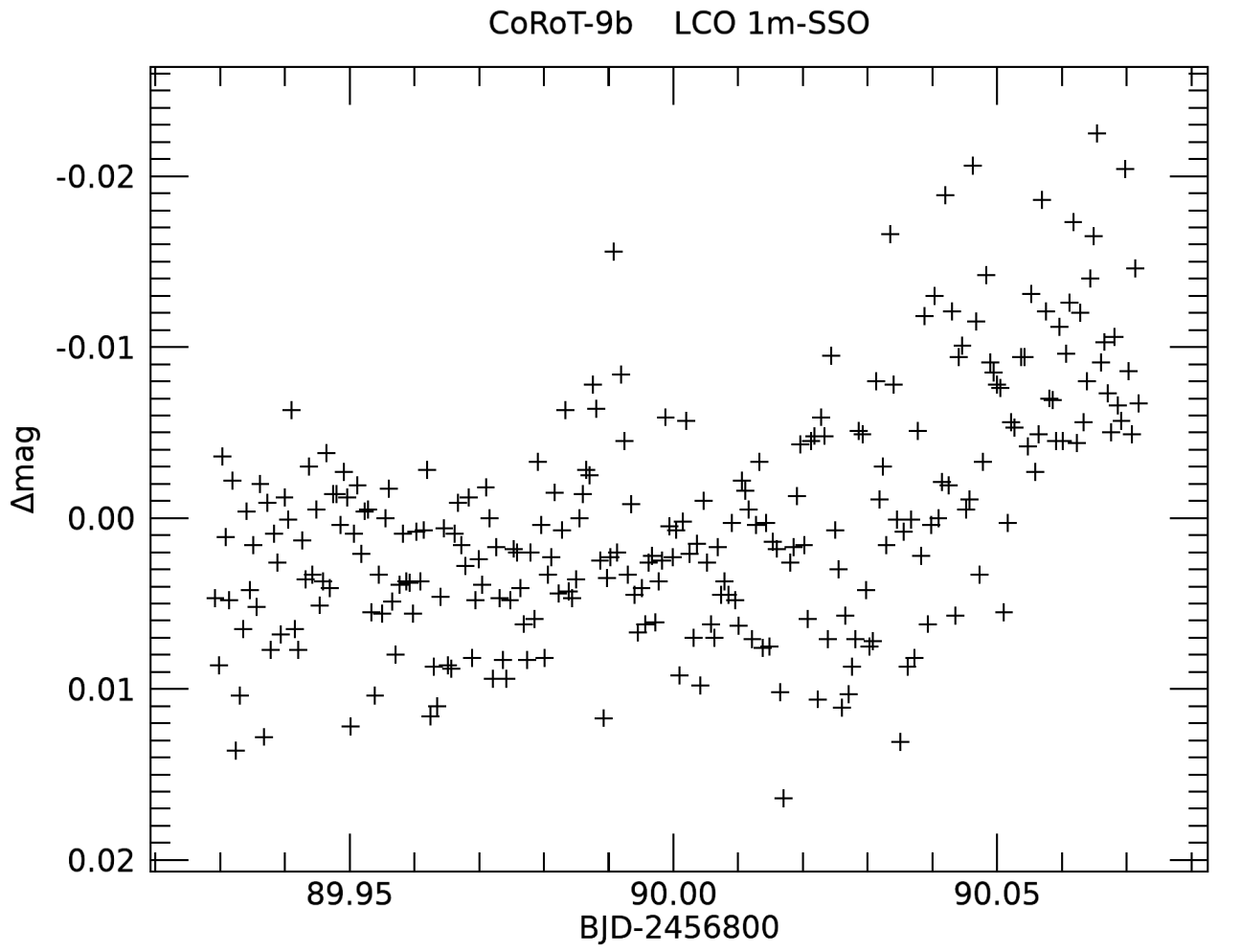

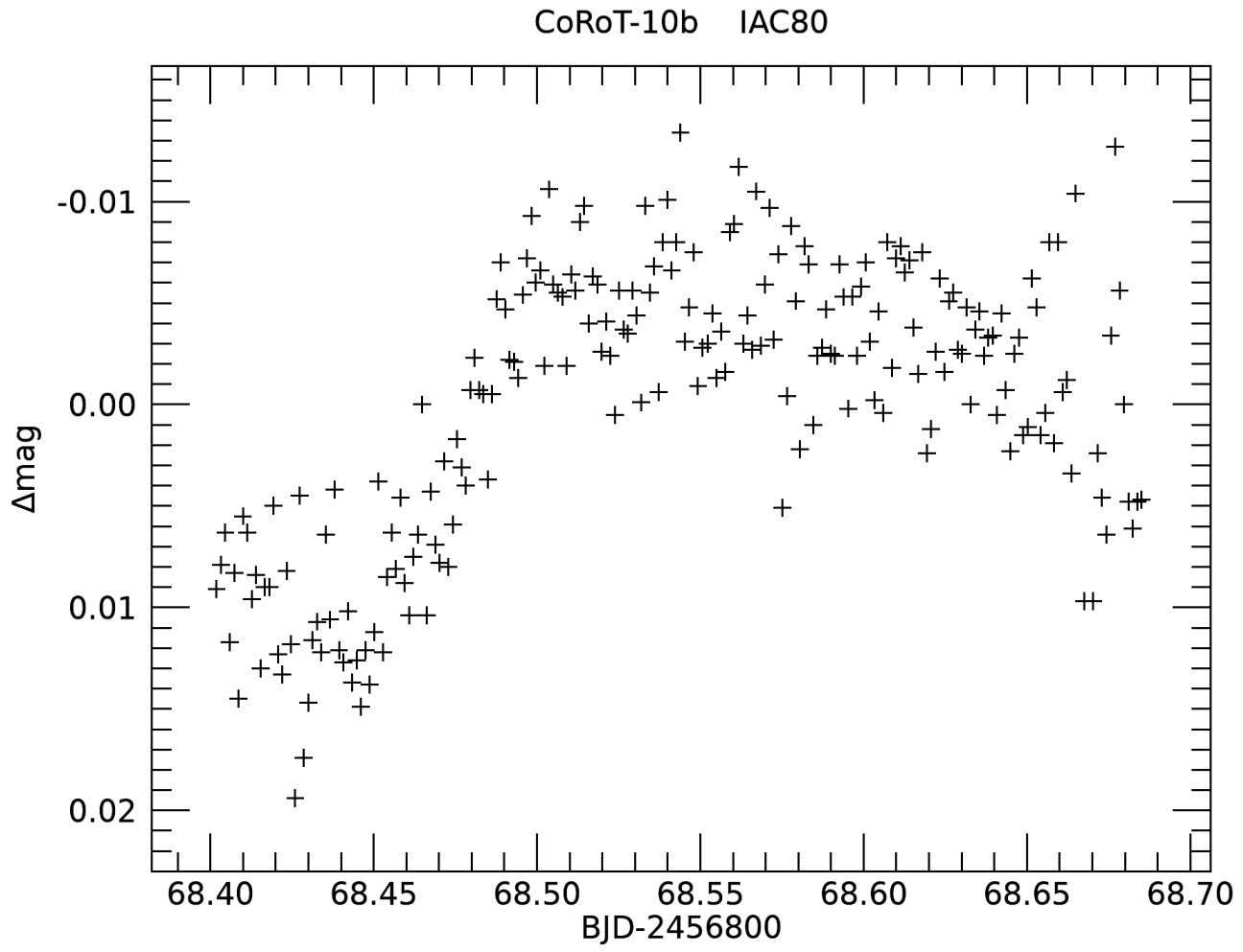

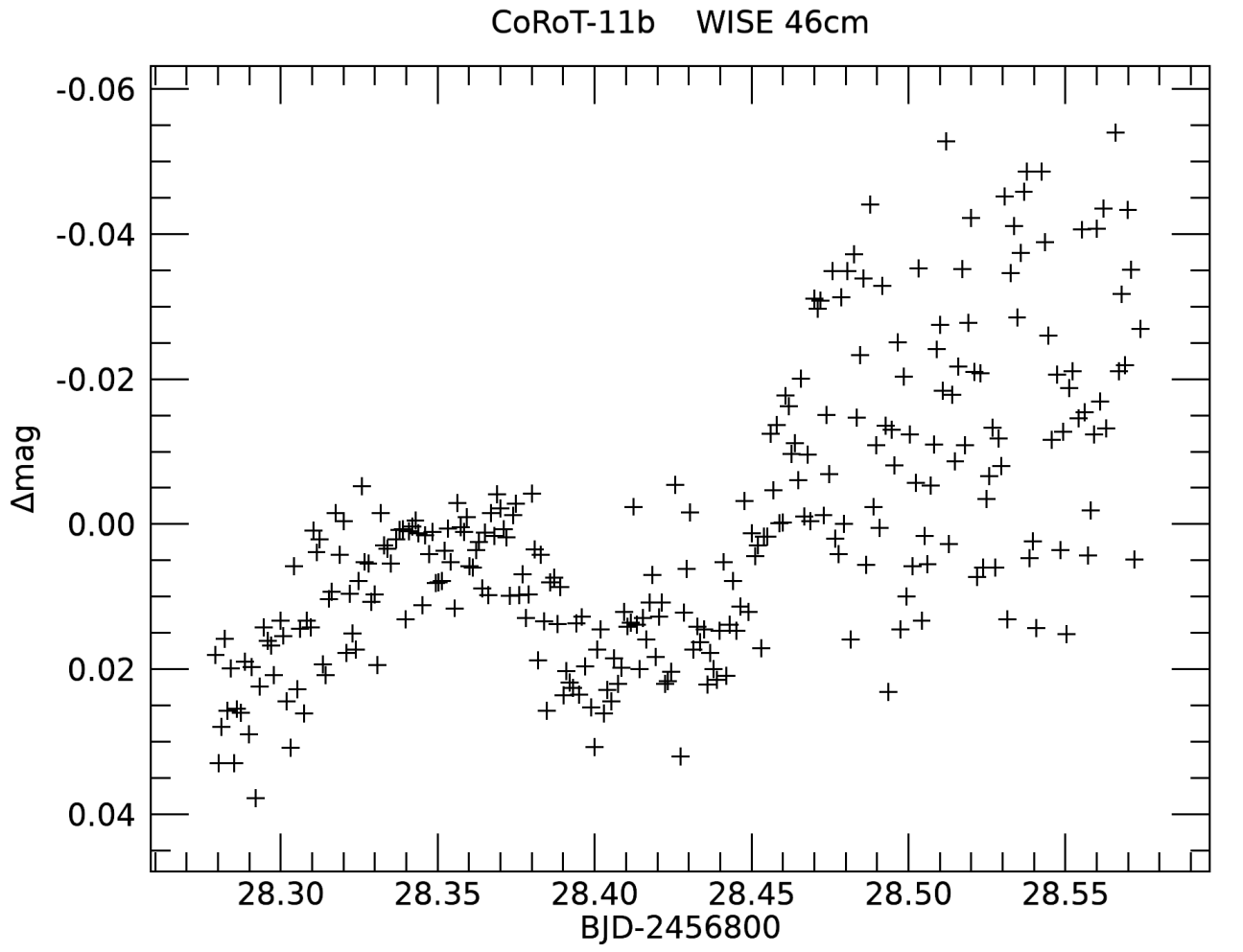

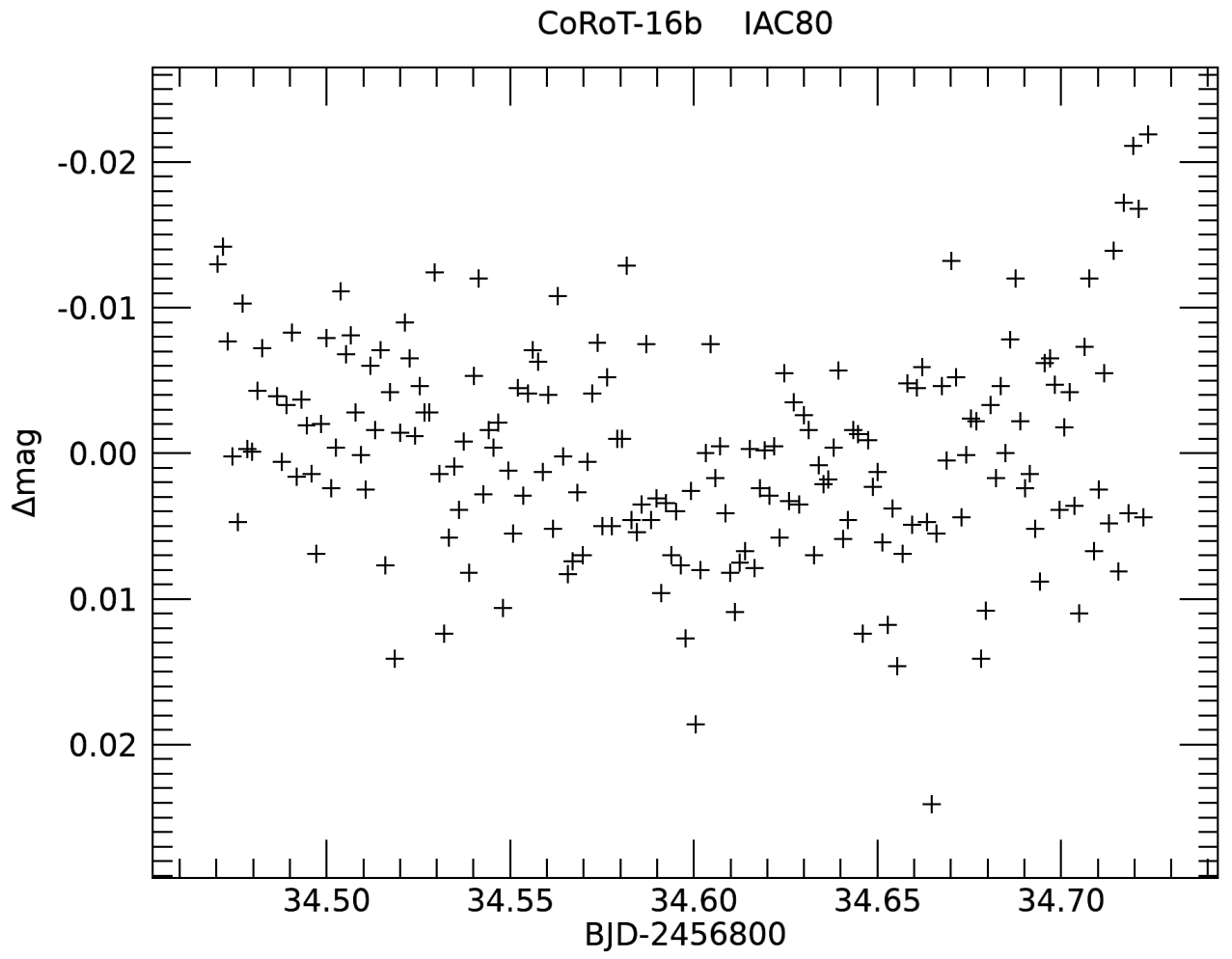
CoRoT-16b IAC80
Δmag
BJD-2456800

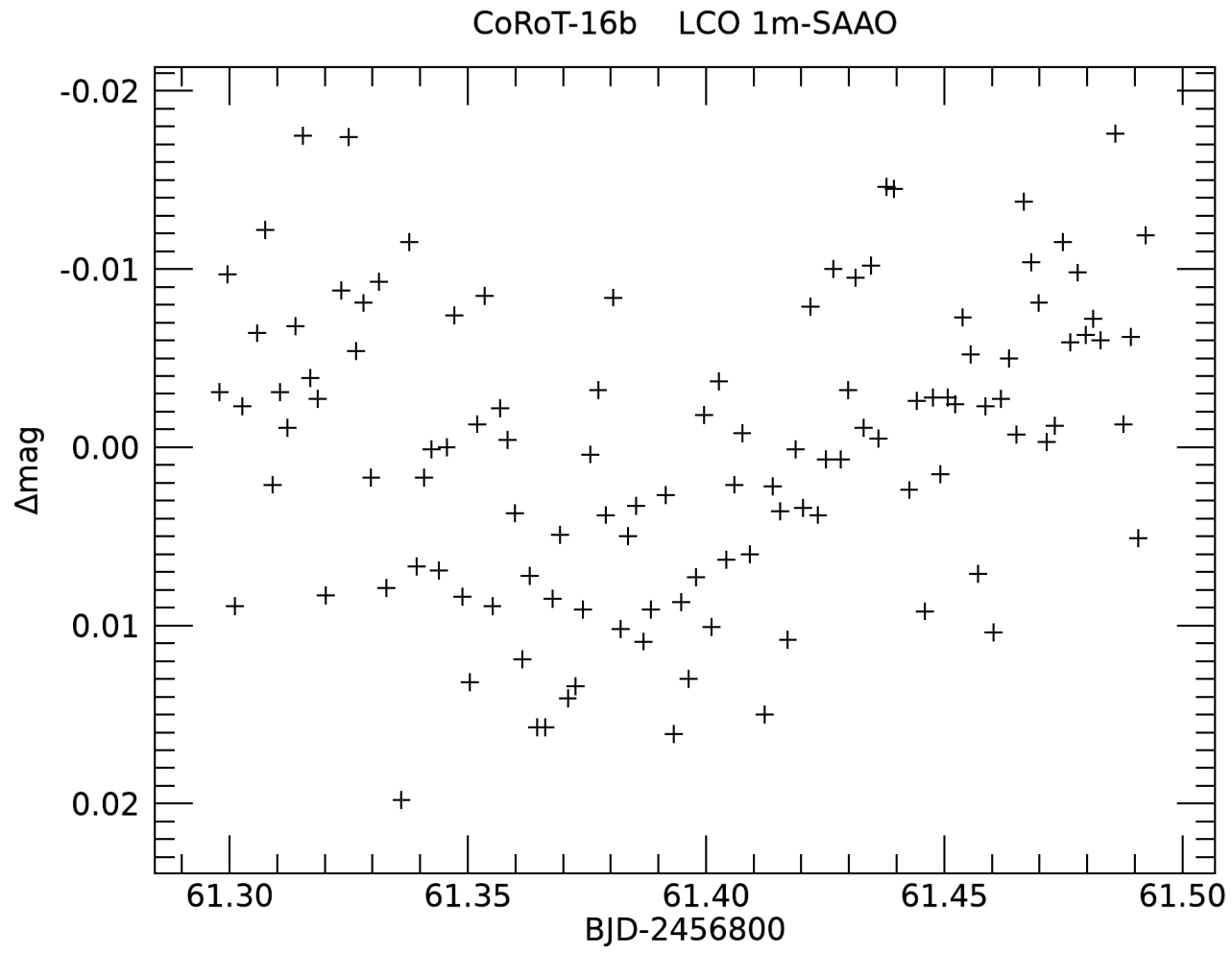
CoRoT-16b LCO 1m-SAAO
Δmag
BJD-2456800

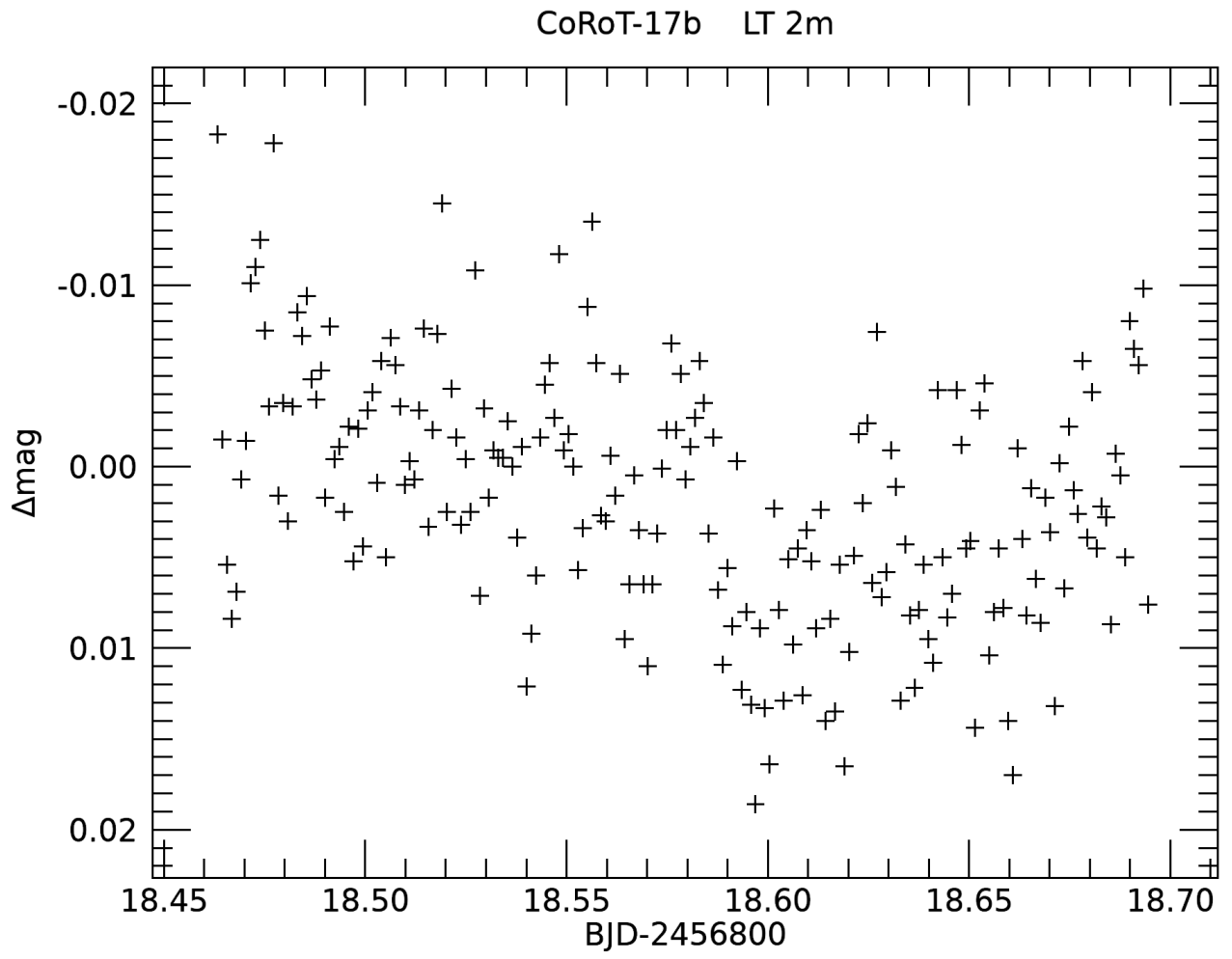
CoRoT-17b LT 2m
Δmag
BJD-2456800

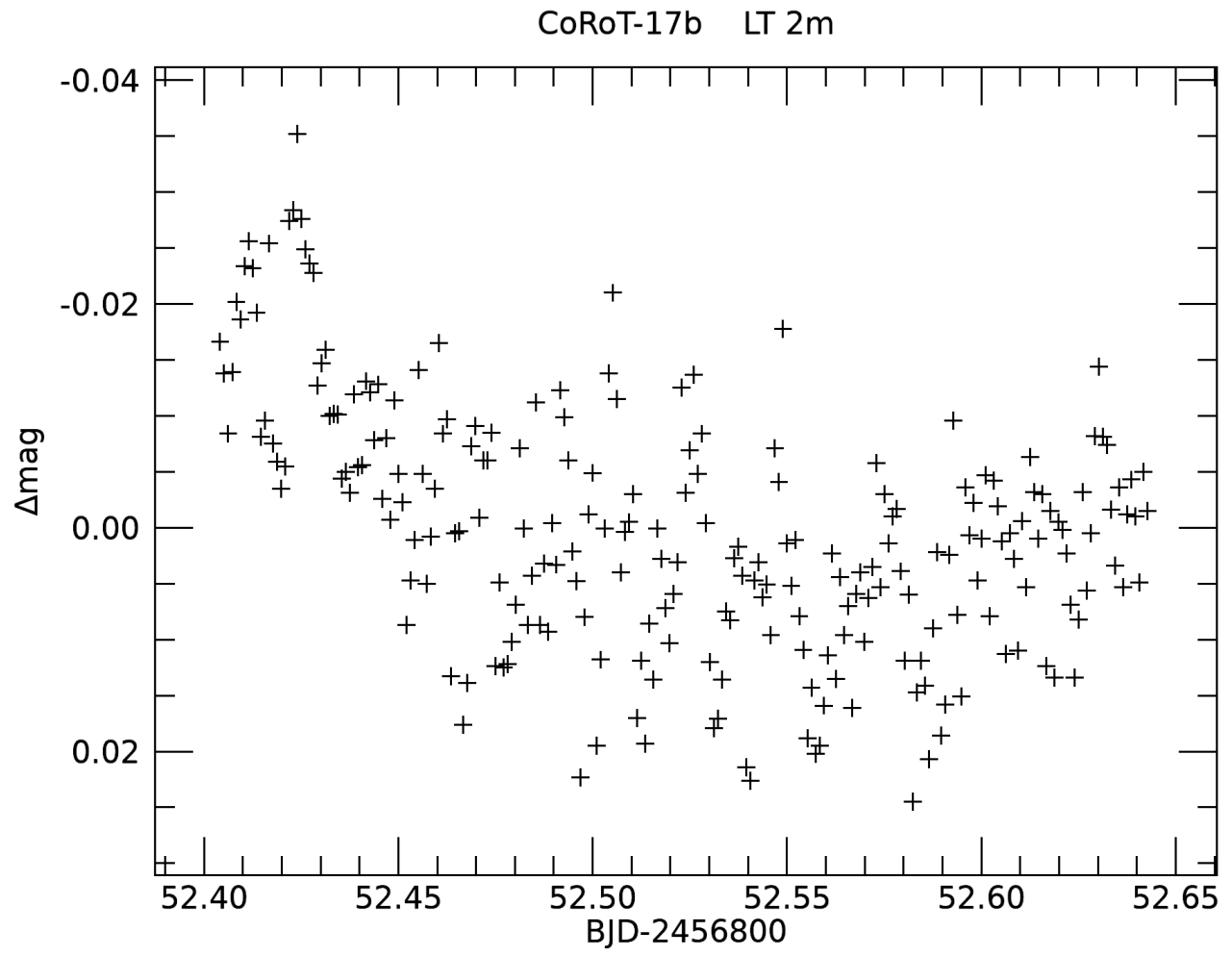
CoRoT-17b LT 2m
Δmag
BJD-2456800

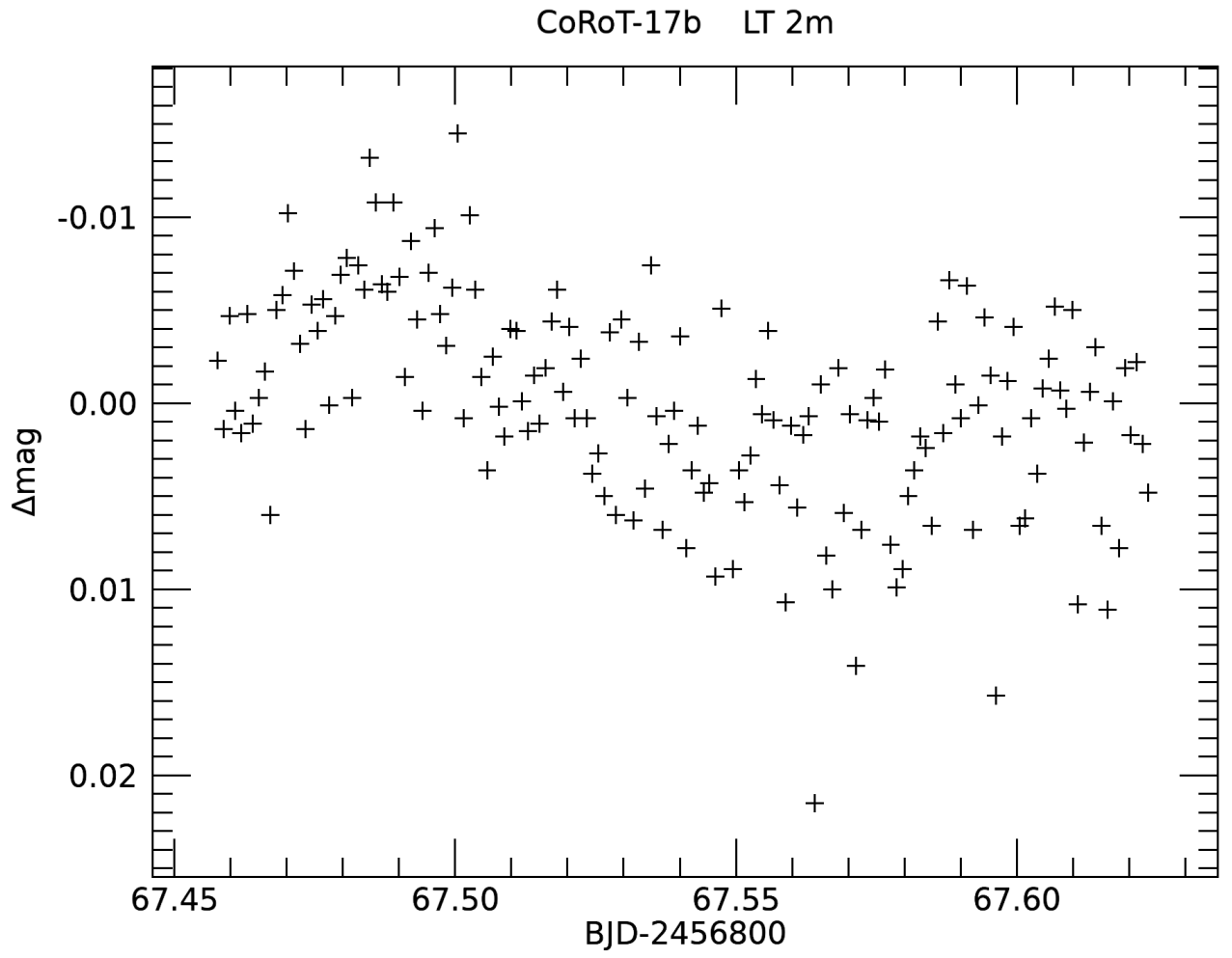
CoRoT-17b LT 2m
Δmag
BJD-2456800

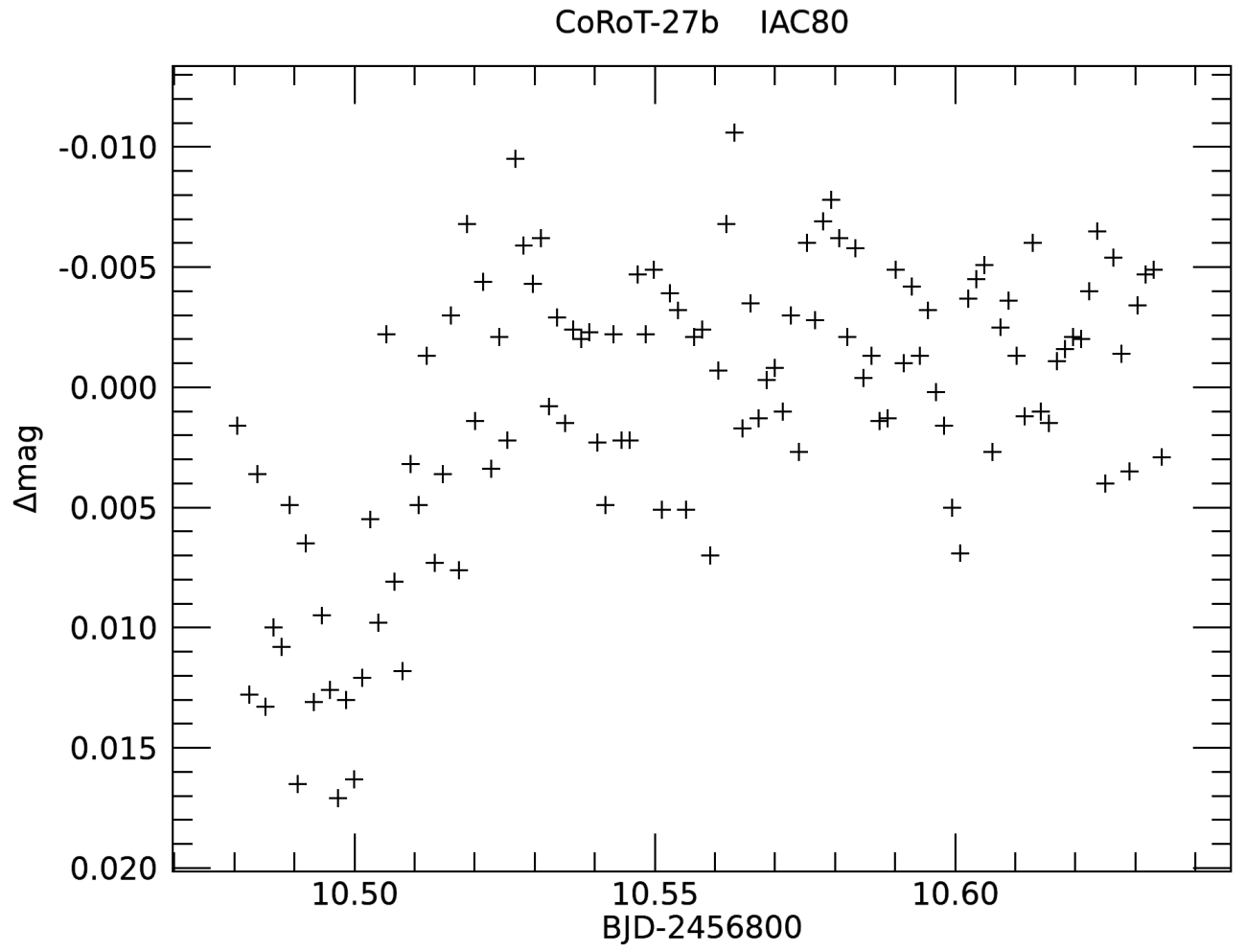
CoRoT-27b IAC80
Δmag
BJD-2456800

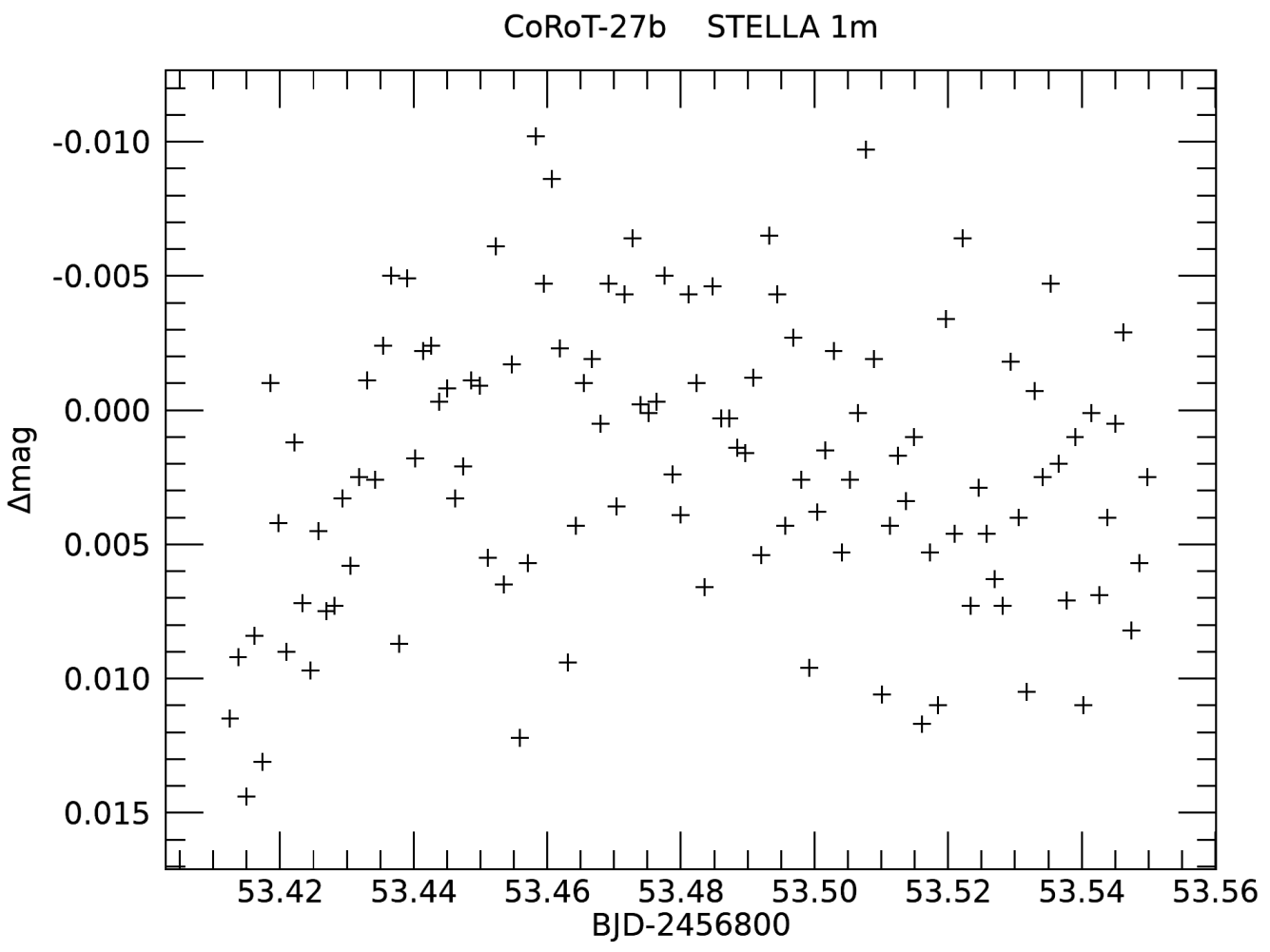
CoRoT-27b STELLA 1m
Δmag
-0.010
-0.005
0.000
0.005
0.010
0.015
53.42 53.44 53.46 53.48 53.50 53.52 53.54 53.56
BJD-2456800

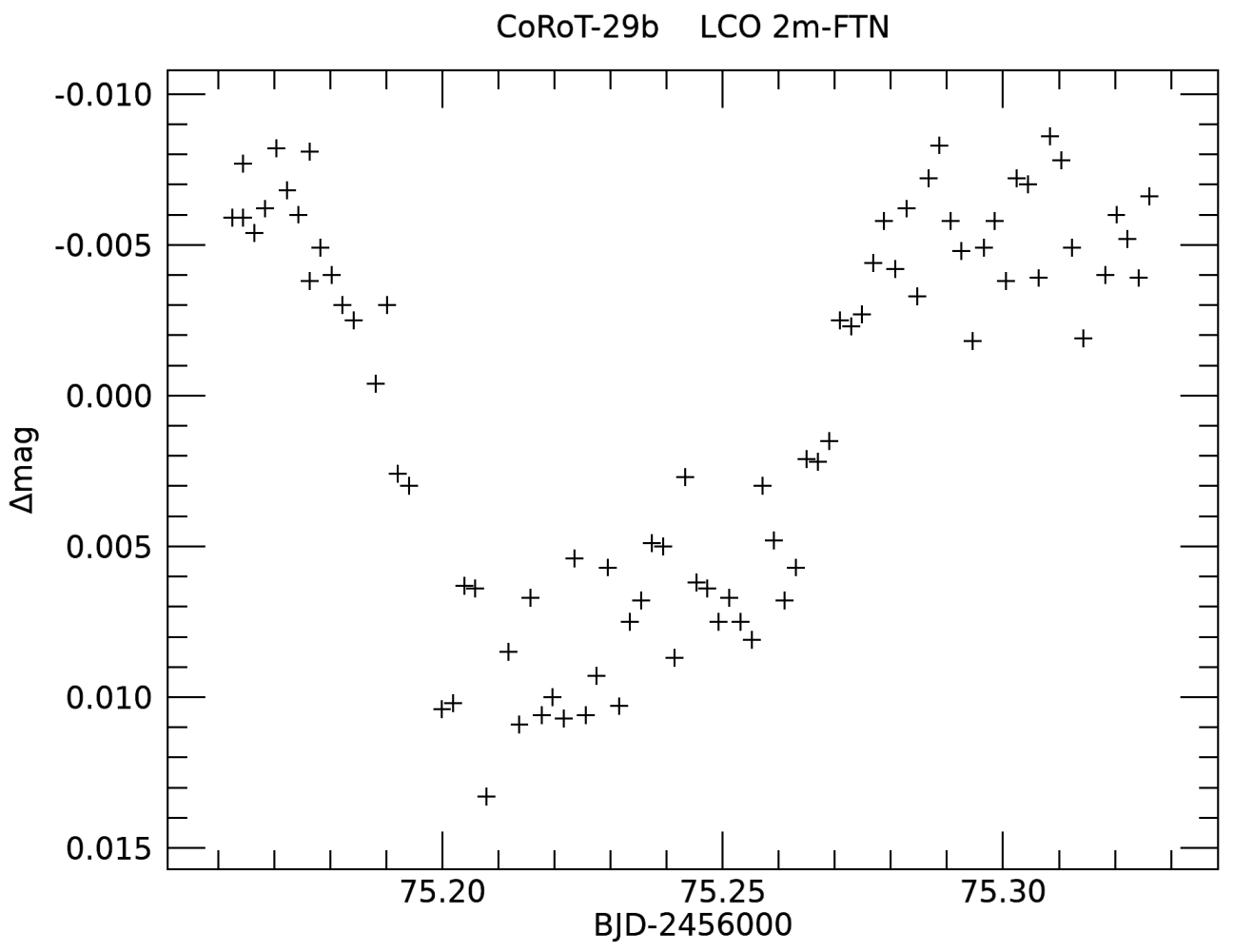

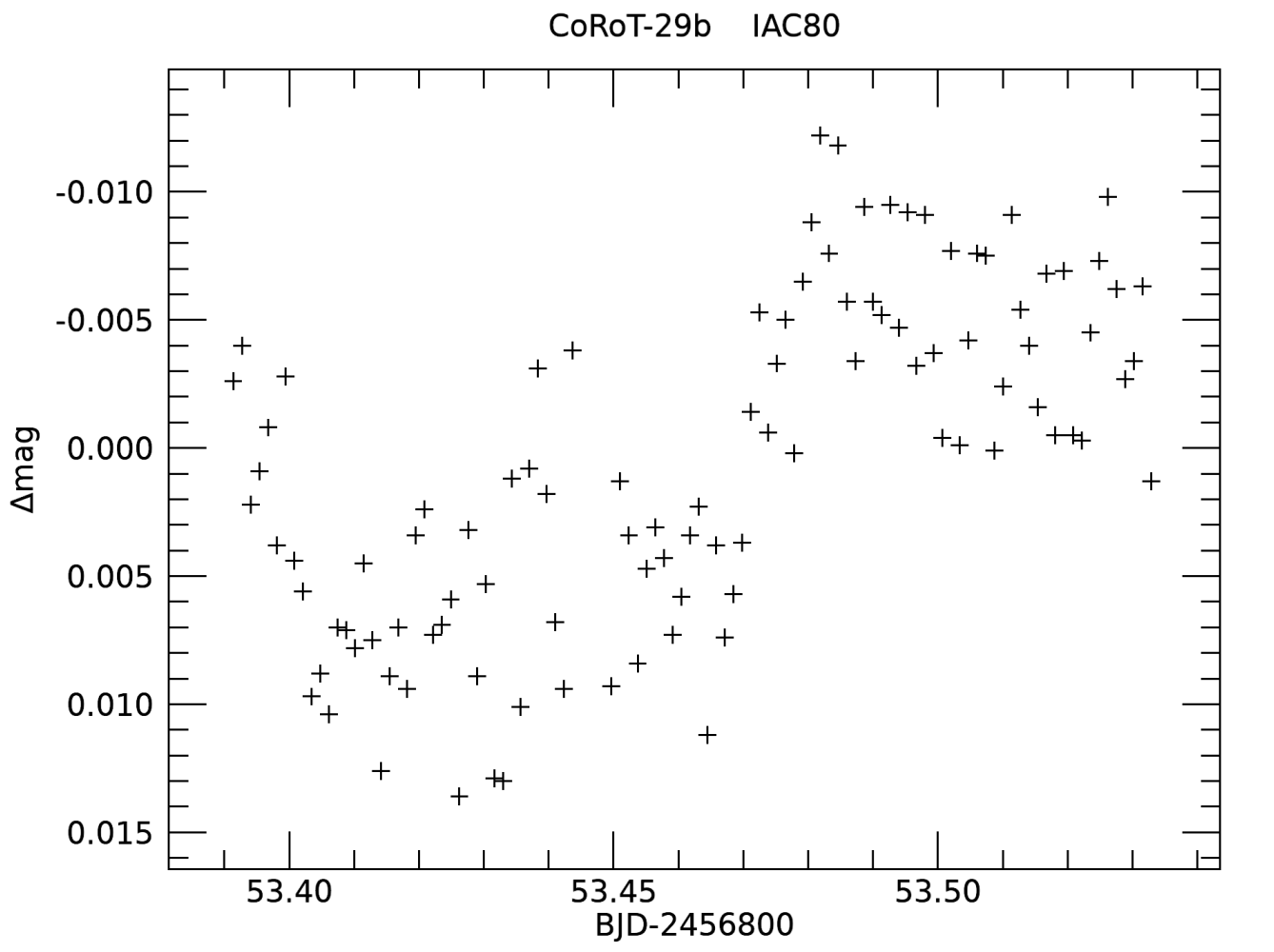

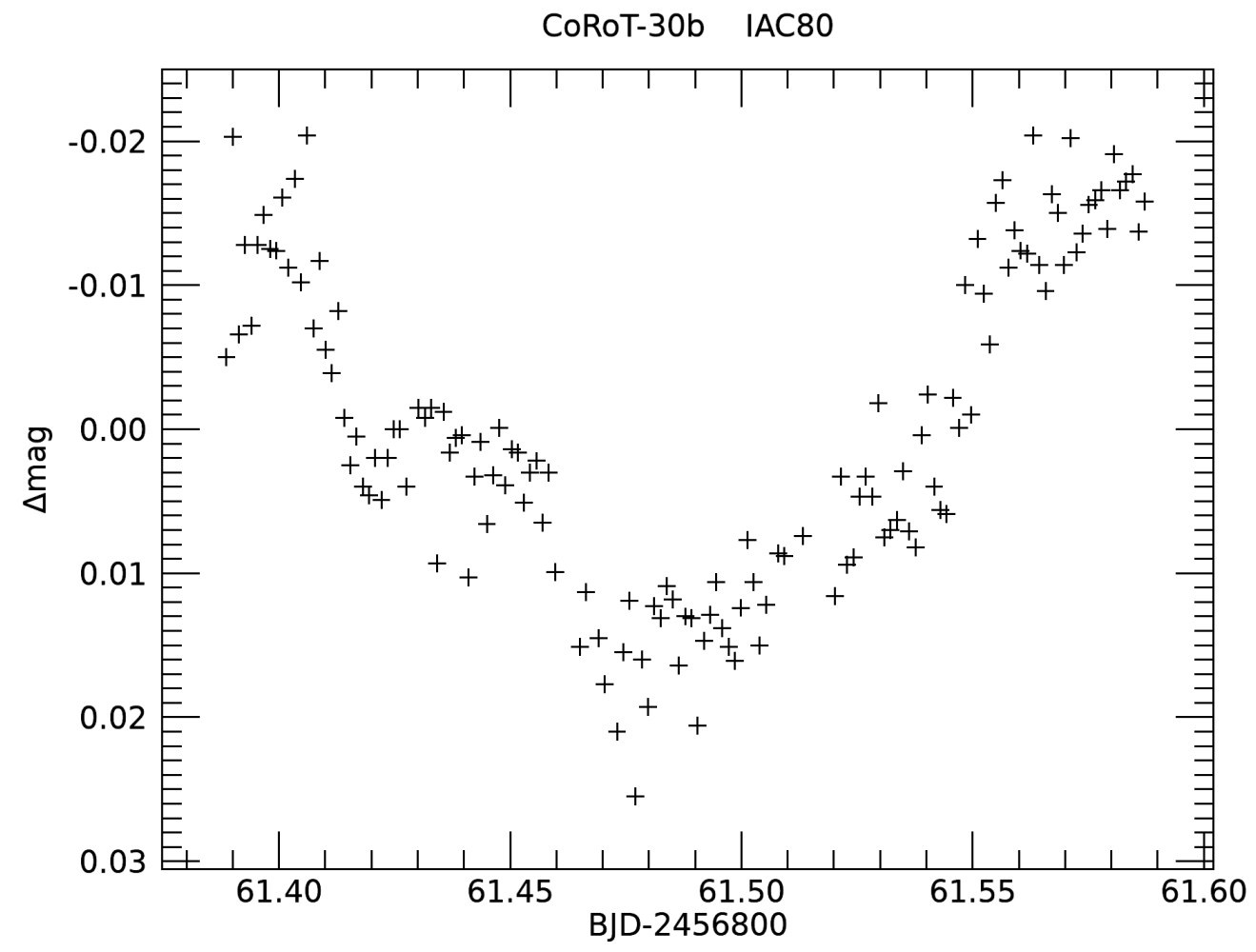

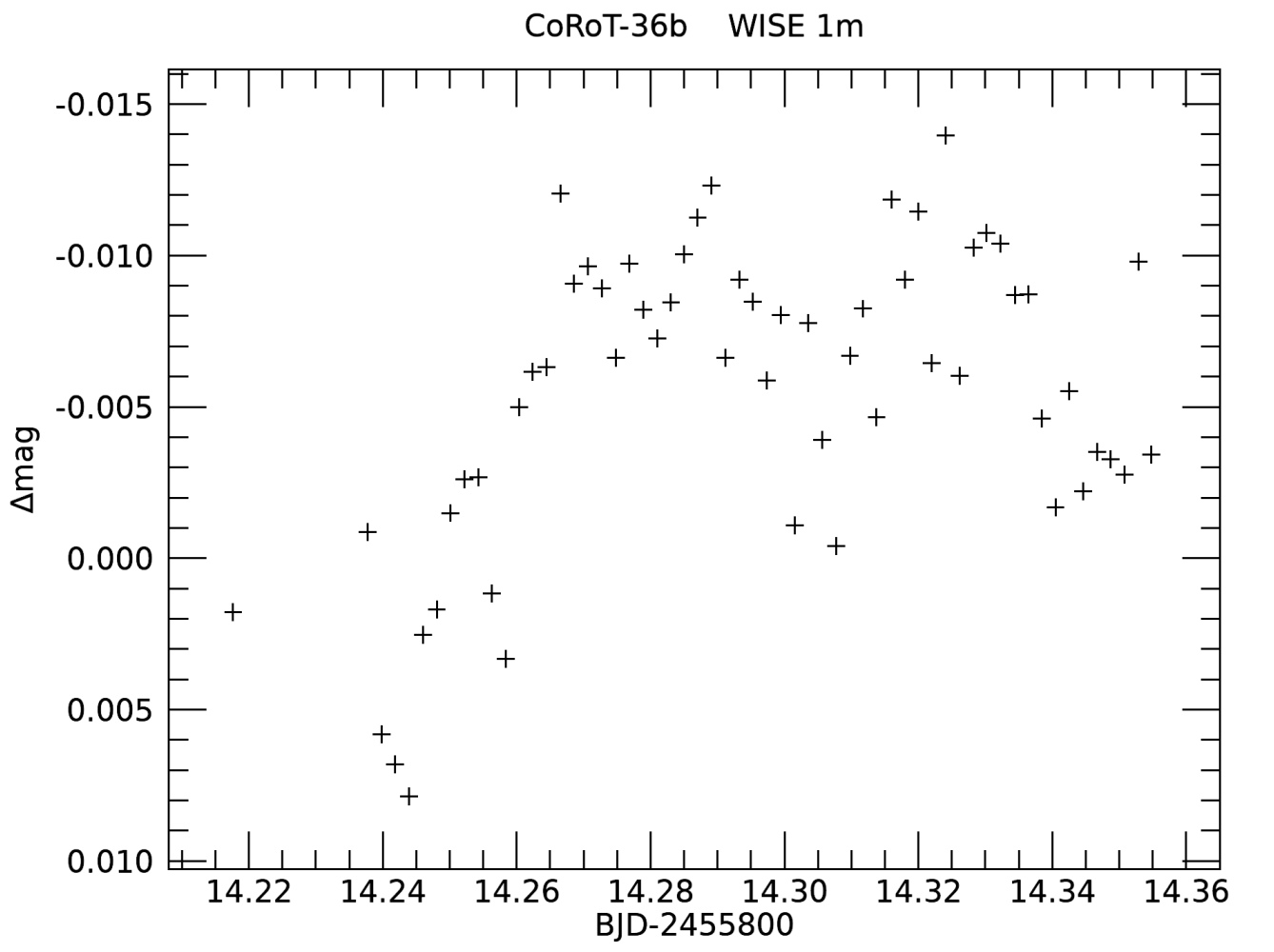
CoRoT-36b WISE 1m
Δmag
-0.015
-0.010
-0.005
0.000
0.005
0.010
14.22 14.24 14.26 14.28 14.30 14.32 14.34 14.36
BJD-2455800

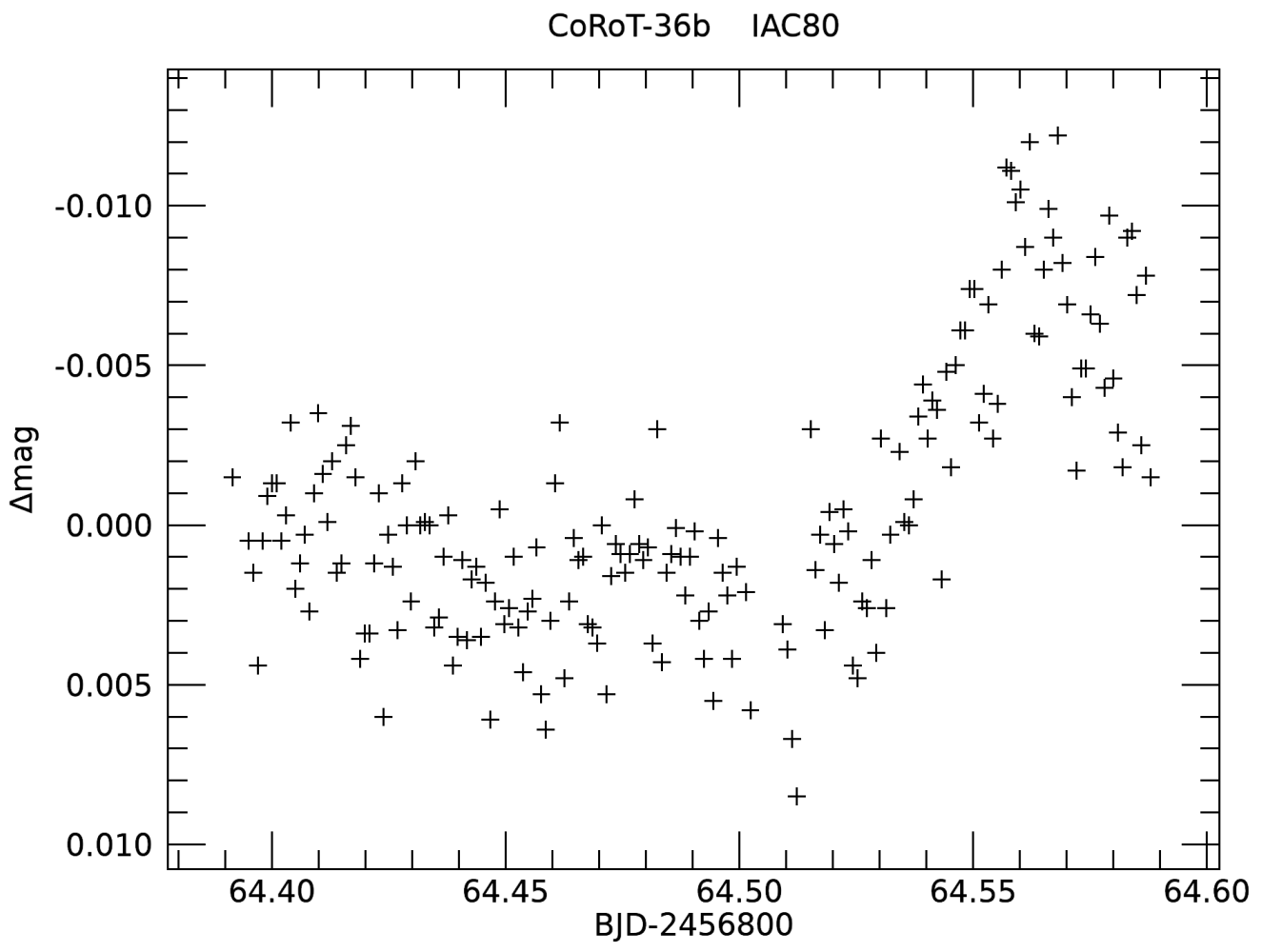

## Appendix B

Transit light curves of CoRoT planets in the galactic anticenter field. They are ordered first by planet number and then by the BJD.

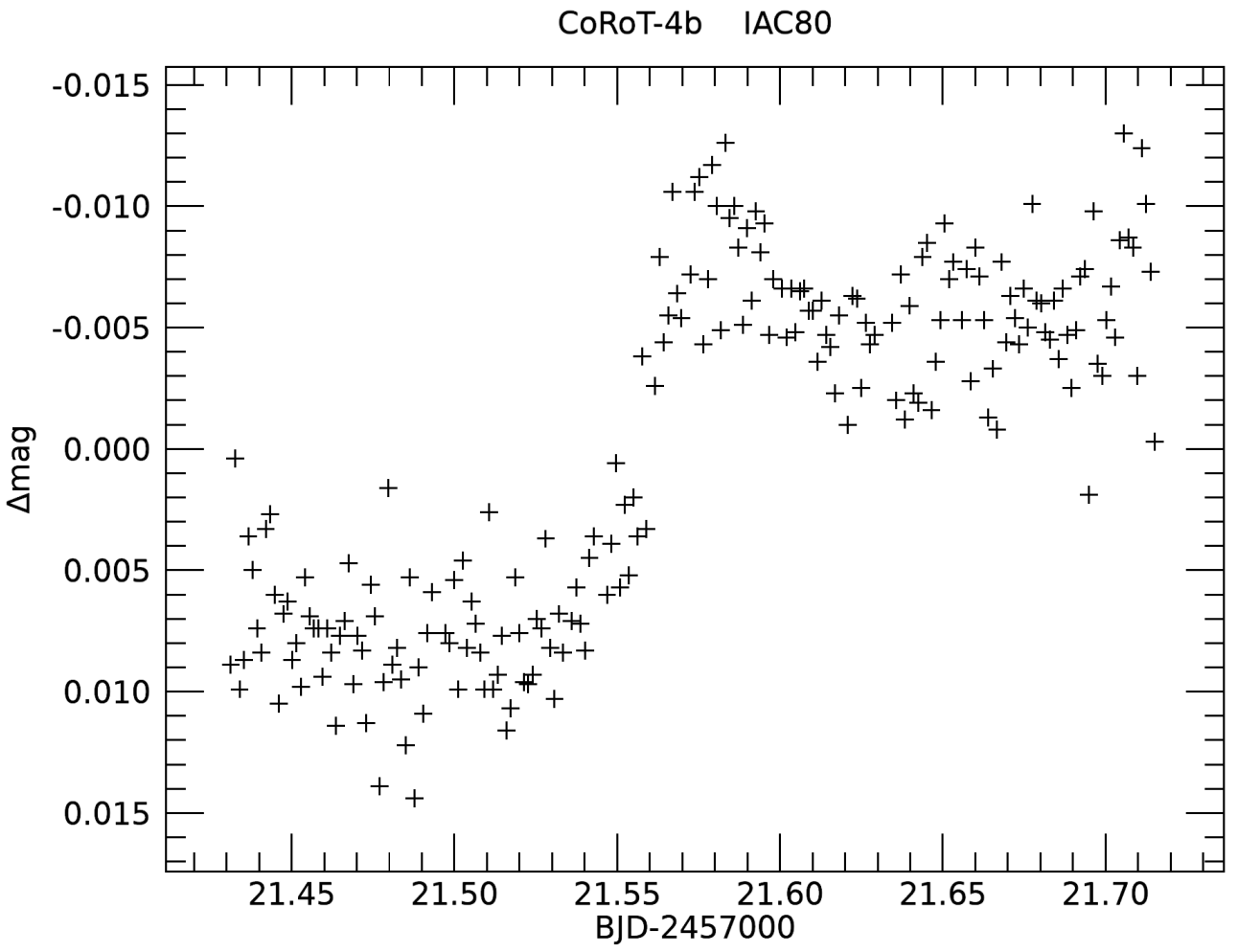

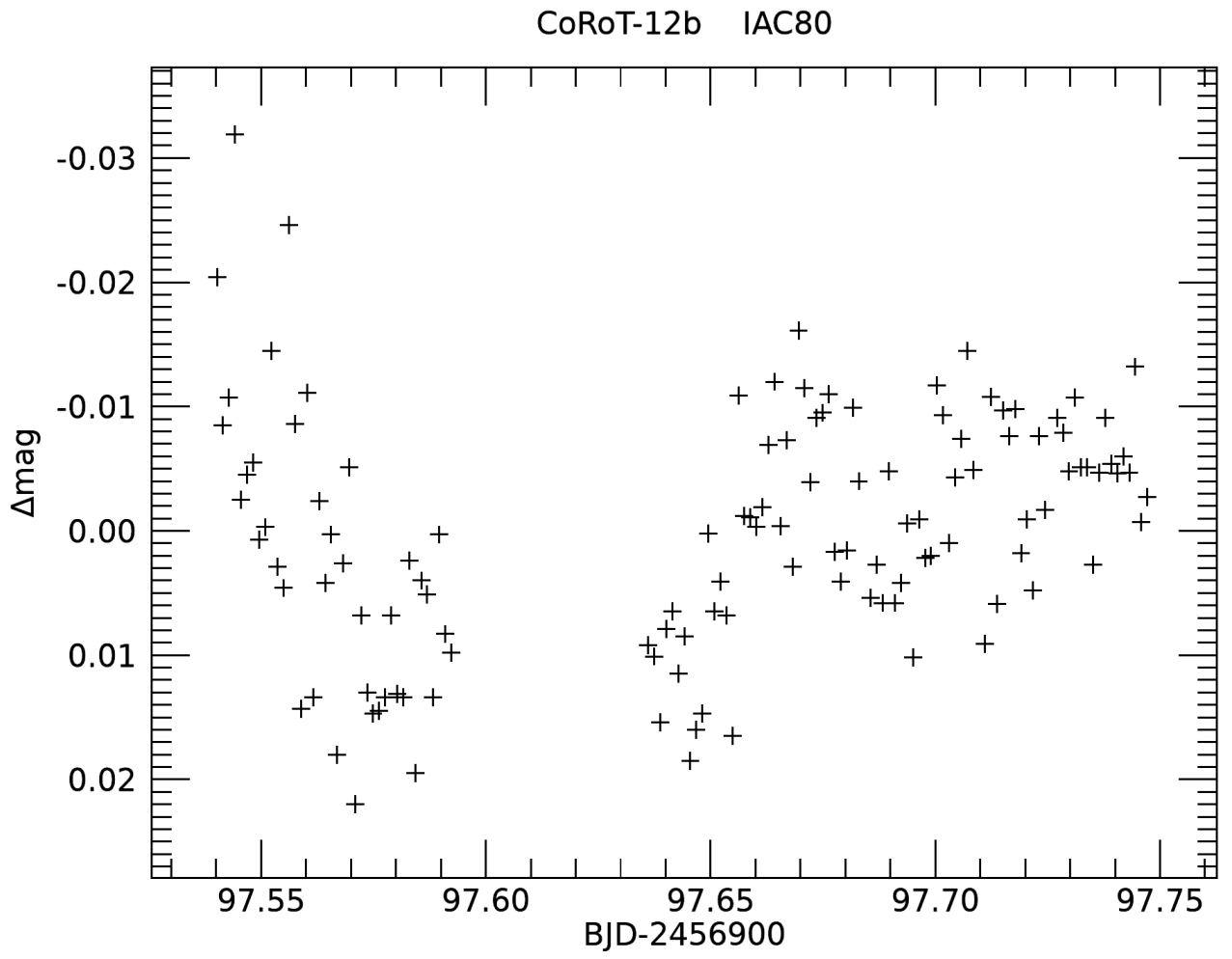

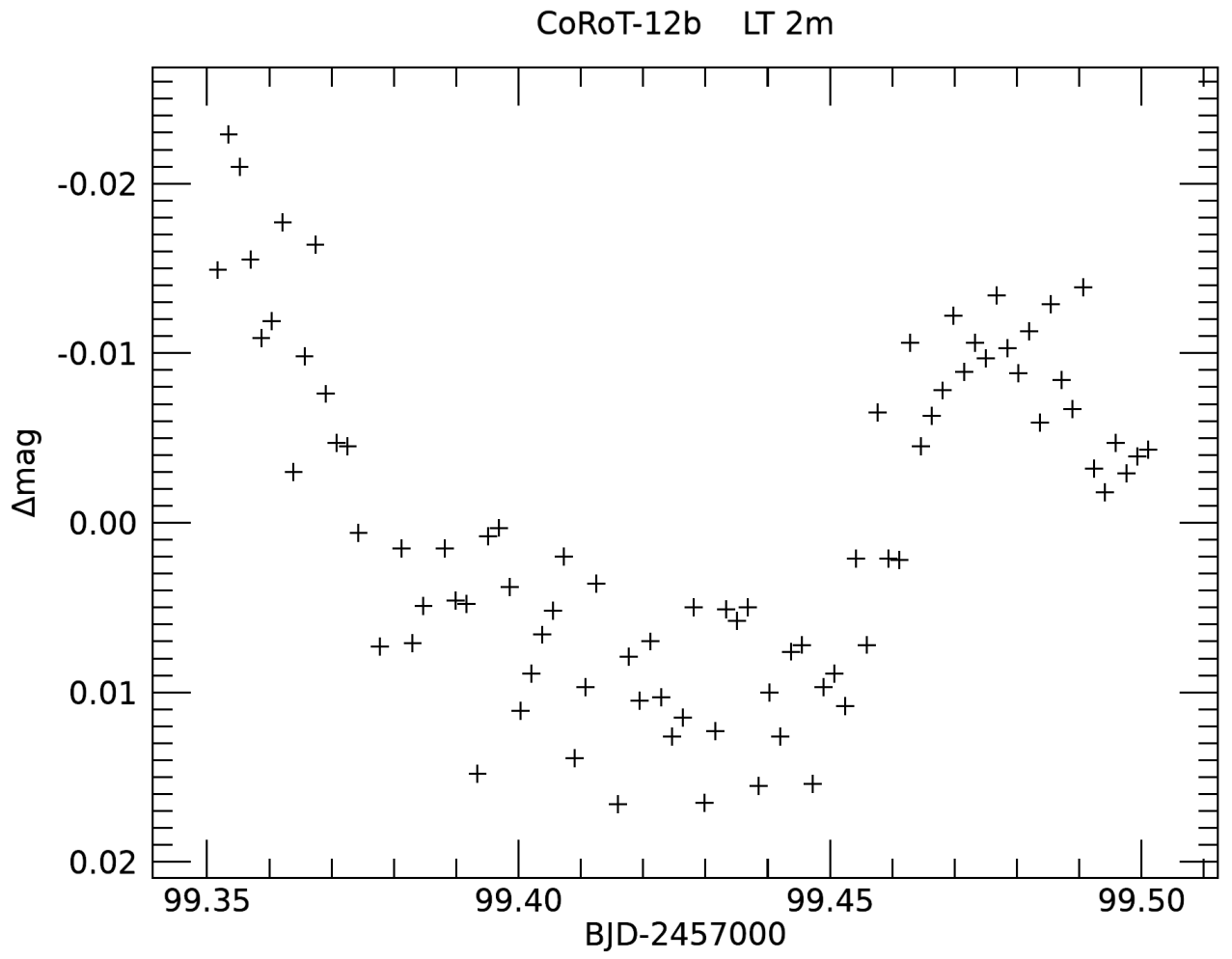

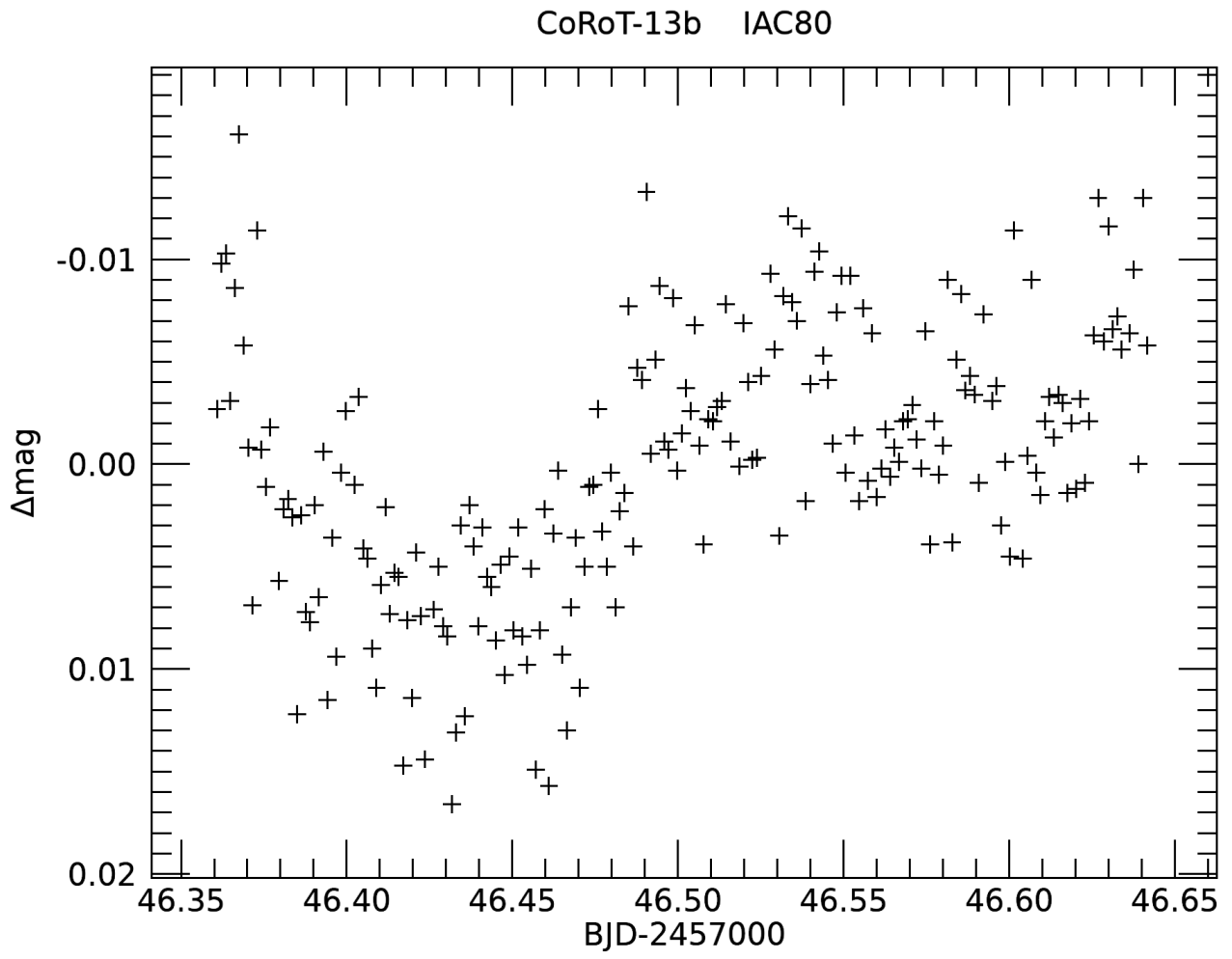

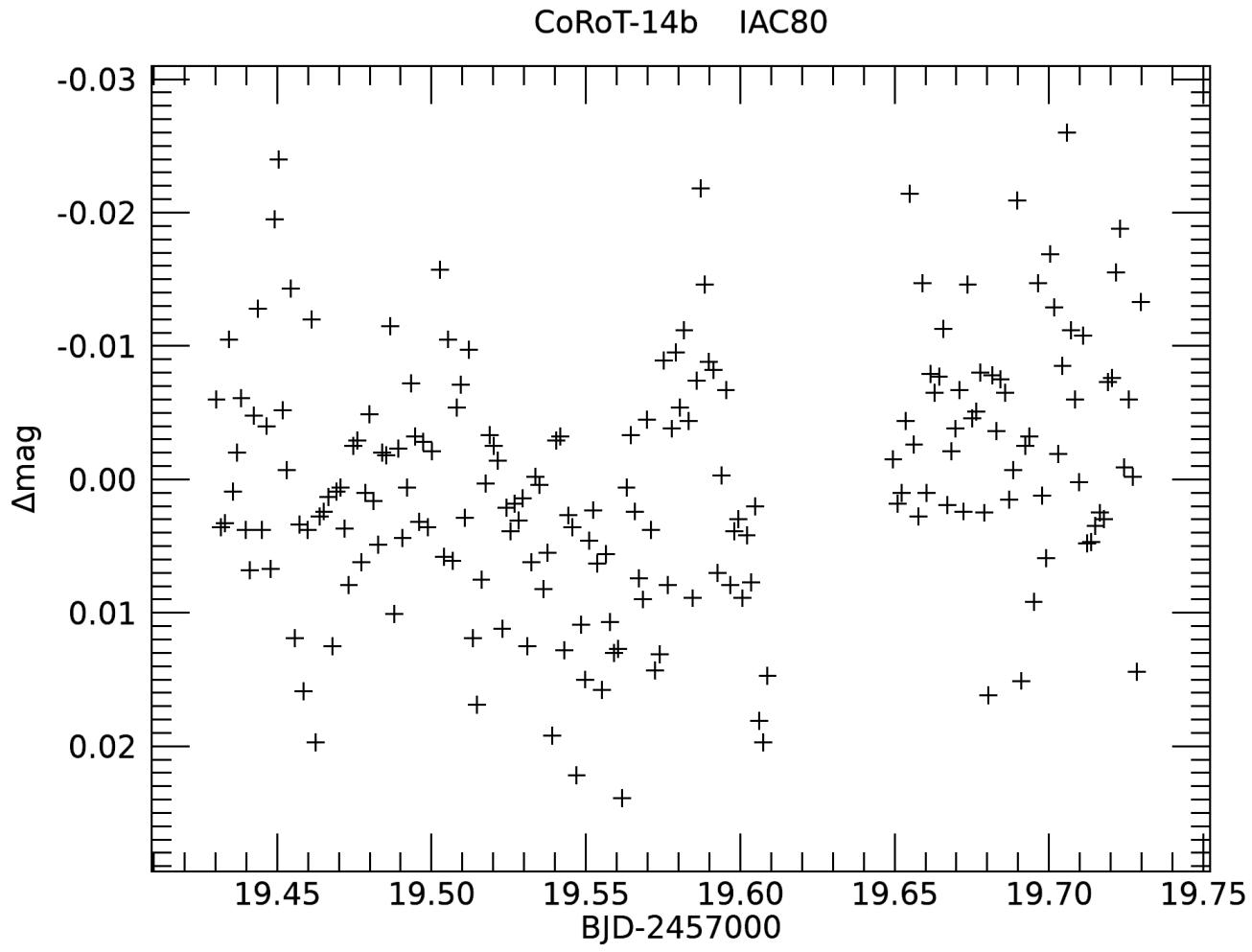

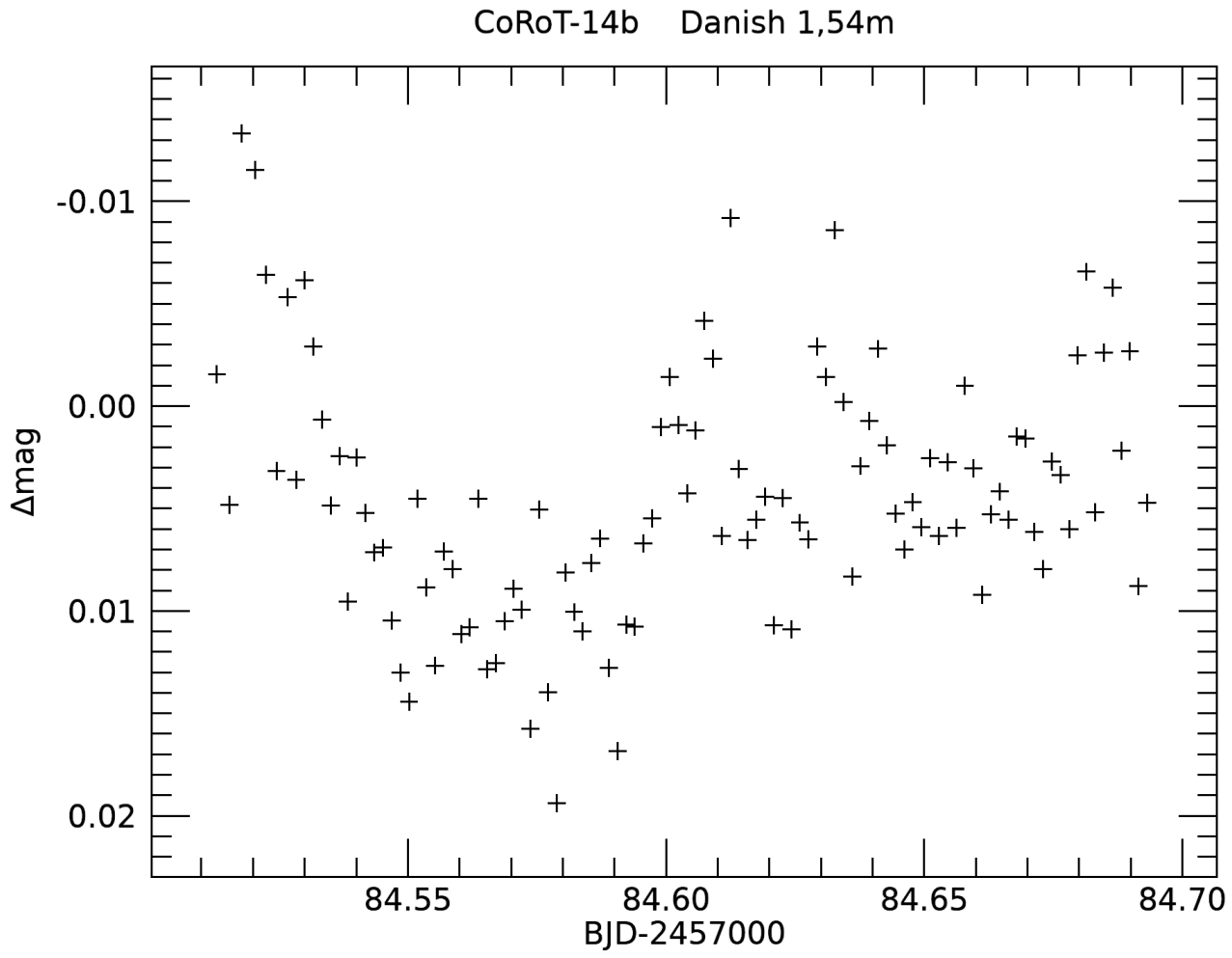

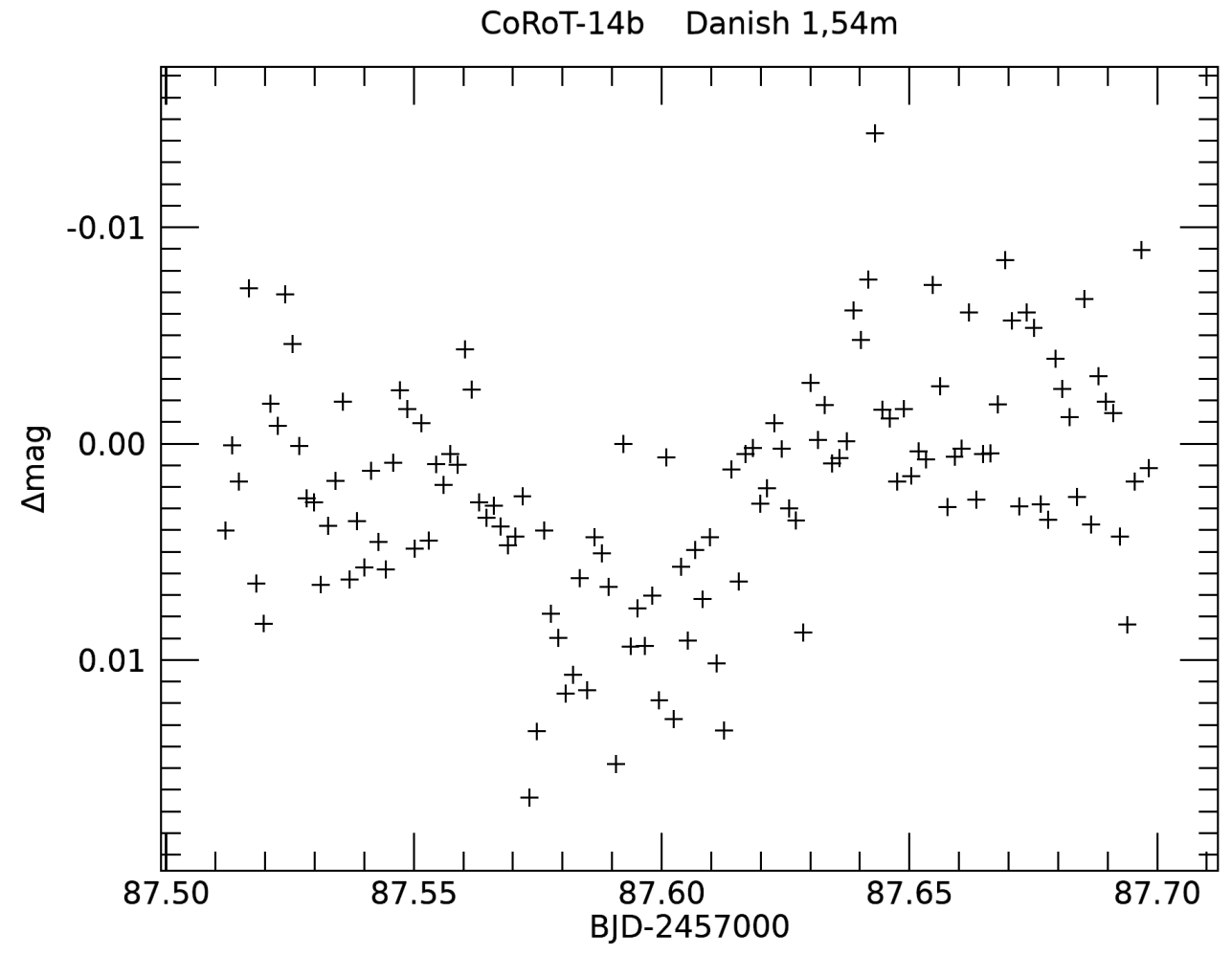
CoRoT-14b Danish 1,54m
Δmag
BJD-2457000

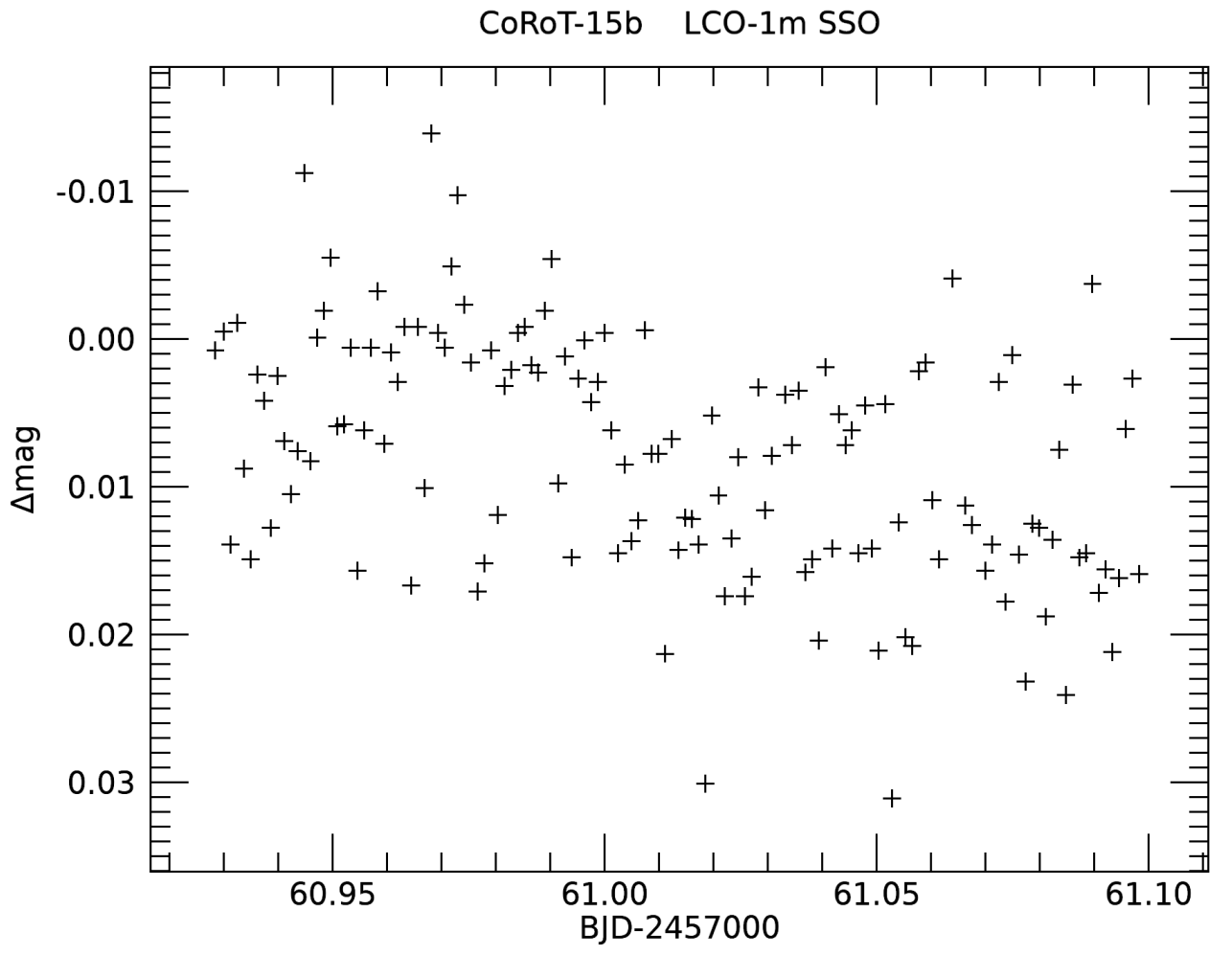
CoRoT-15b LCO-1m SSO
Δmag
BJD-2457000

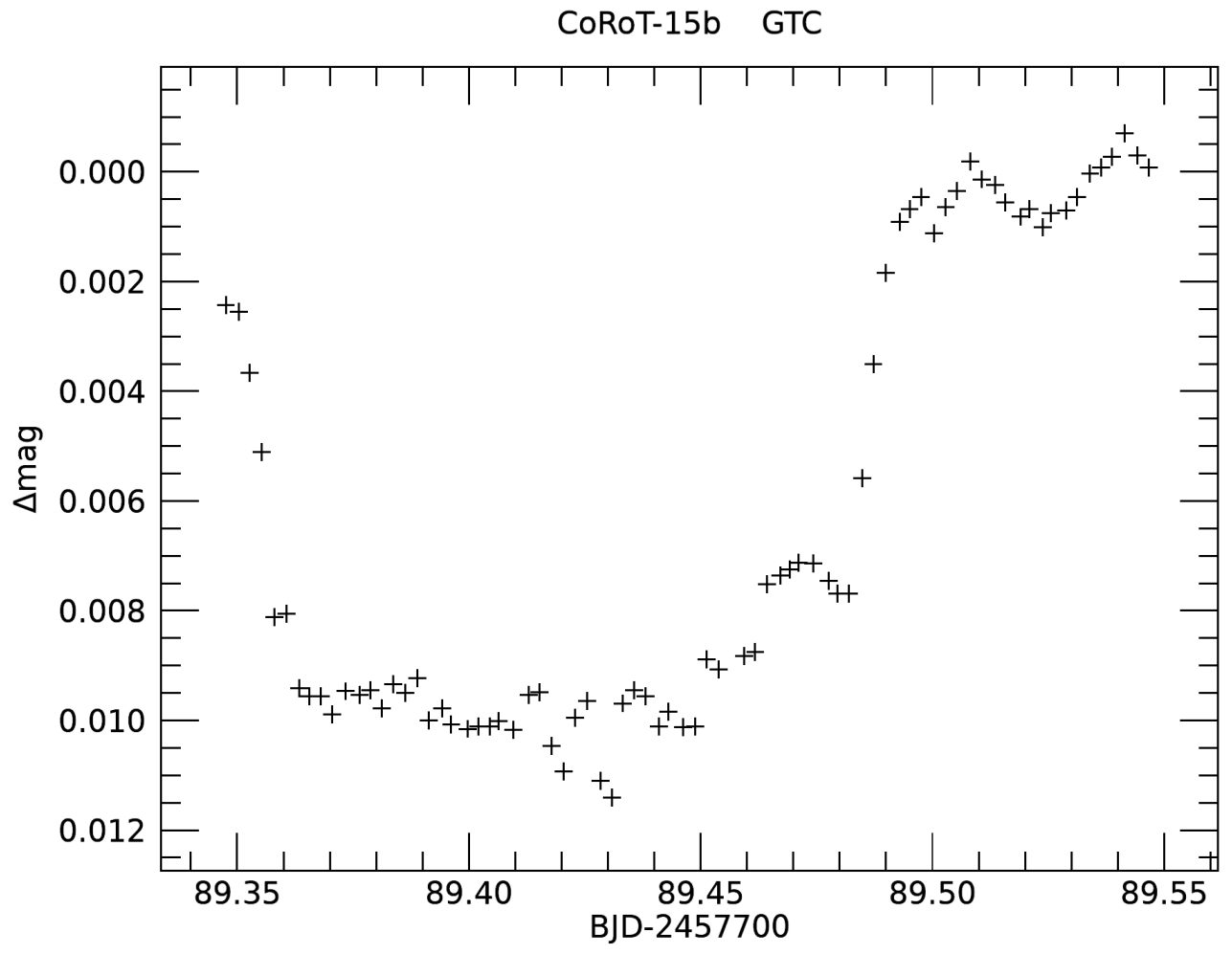
CoRoT-15b GTC
Δmag
BJD-2457700

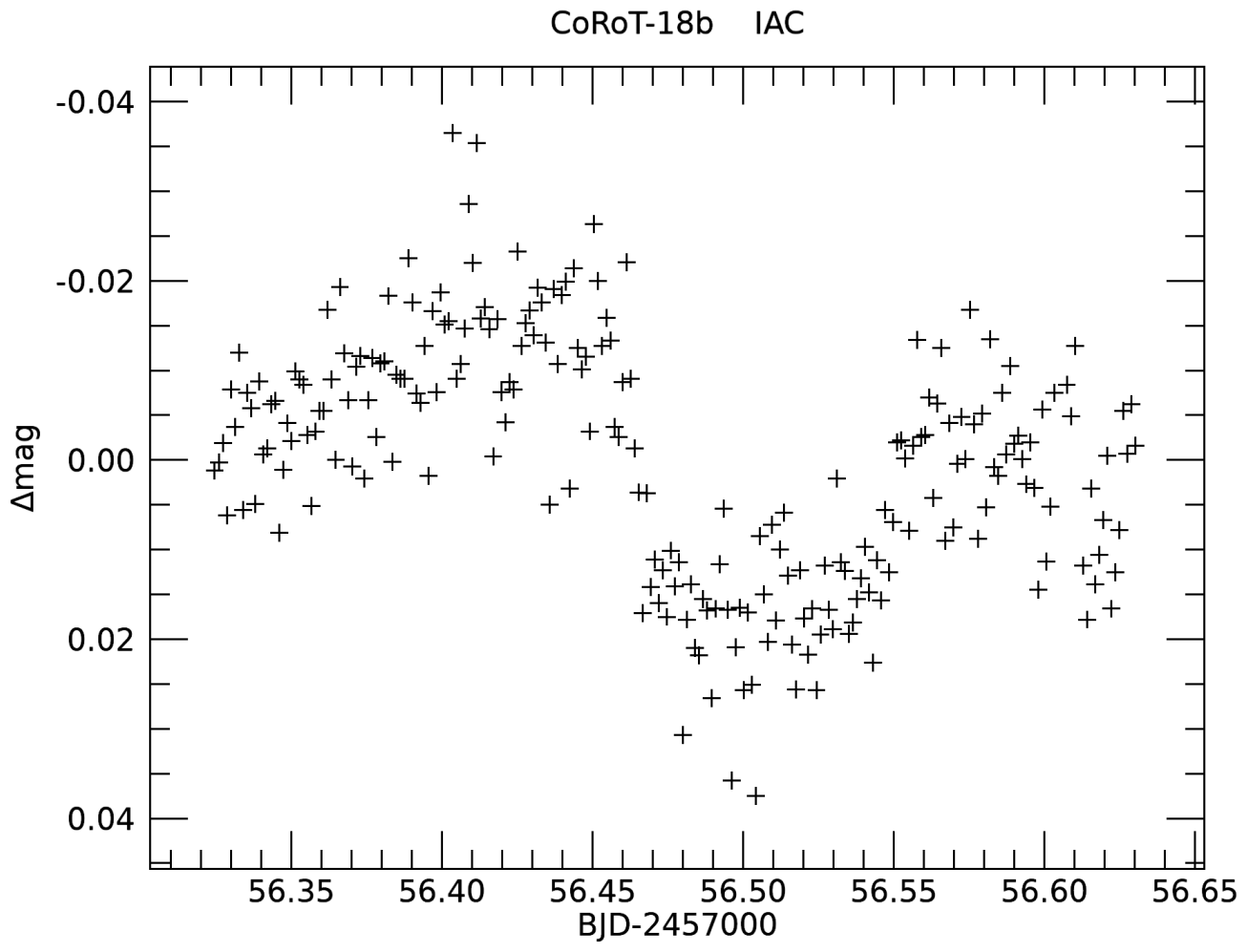
CoRoT-18b IAC
Δmag
BJD-2457000

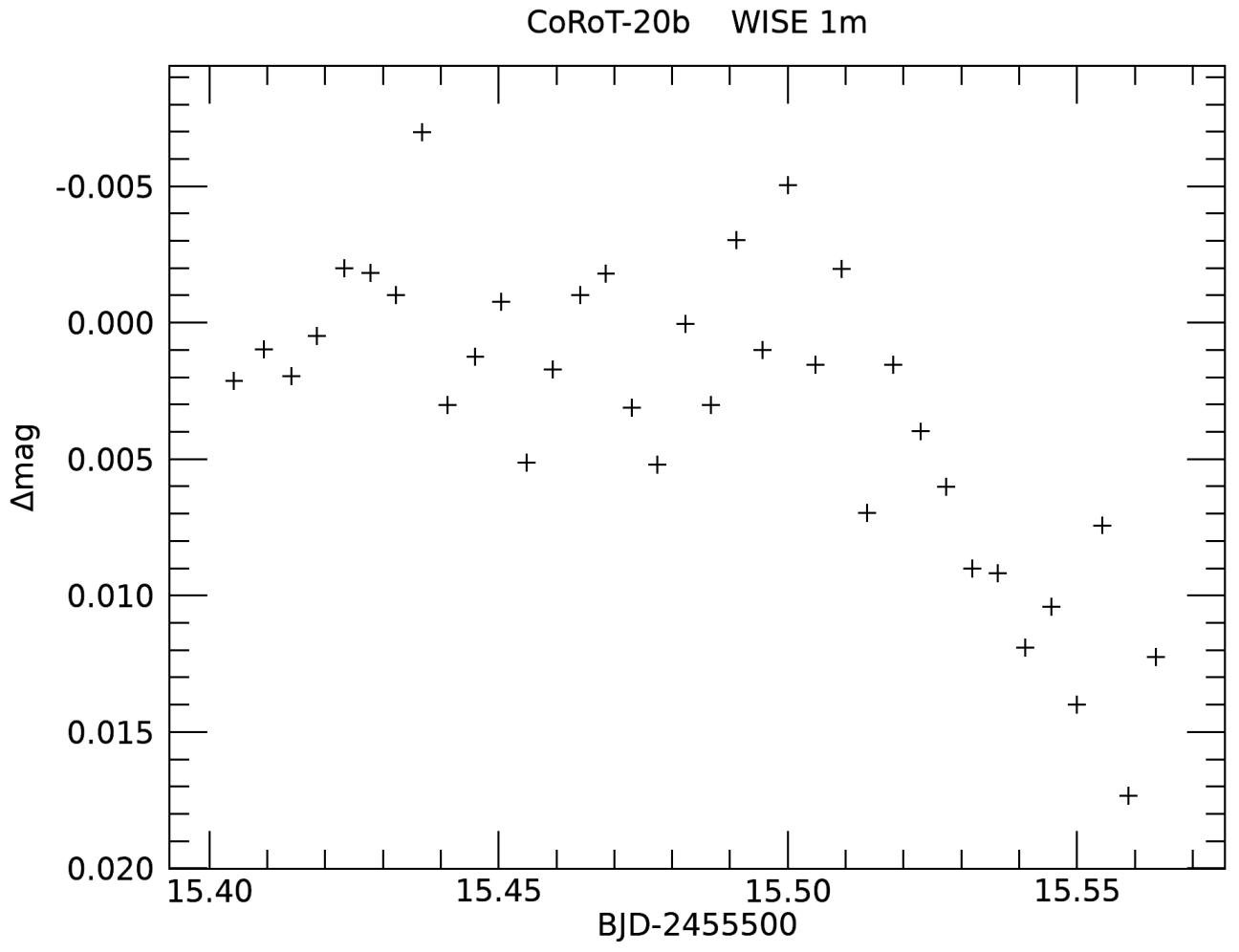
CoRoT-20b WISE 1m
Δmag
BJD-2455500

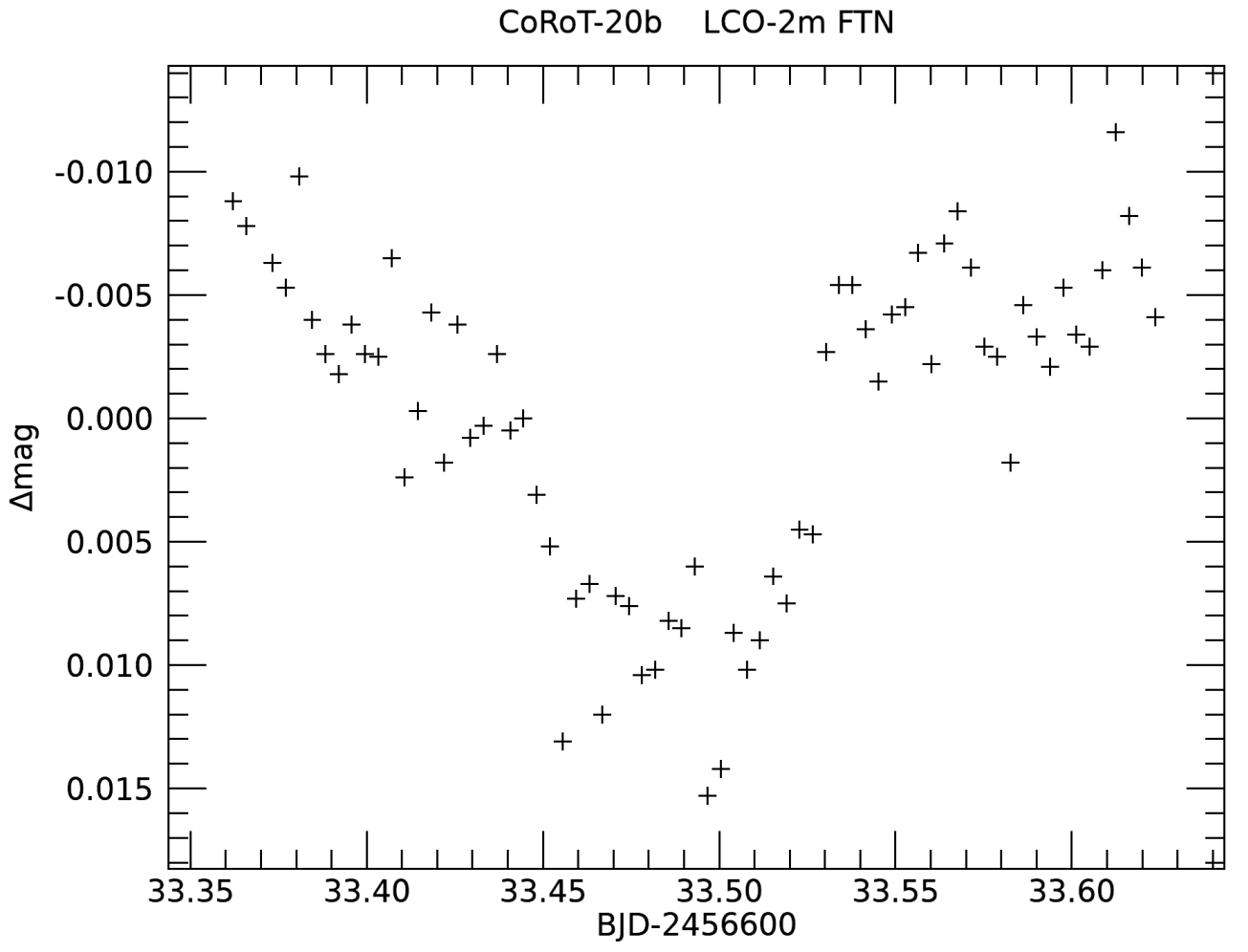
CoRoT-20b LCO-2m FTN
Δmag
BJD-2456600

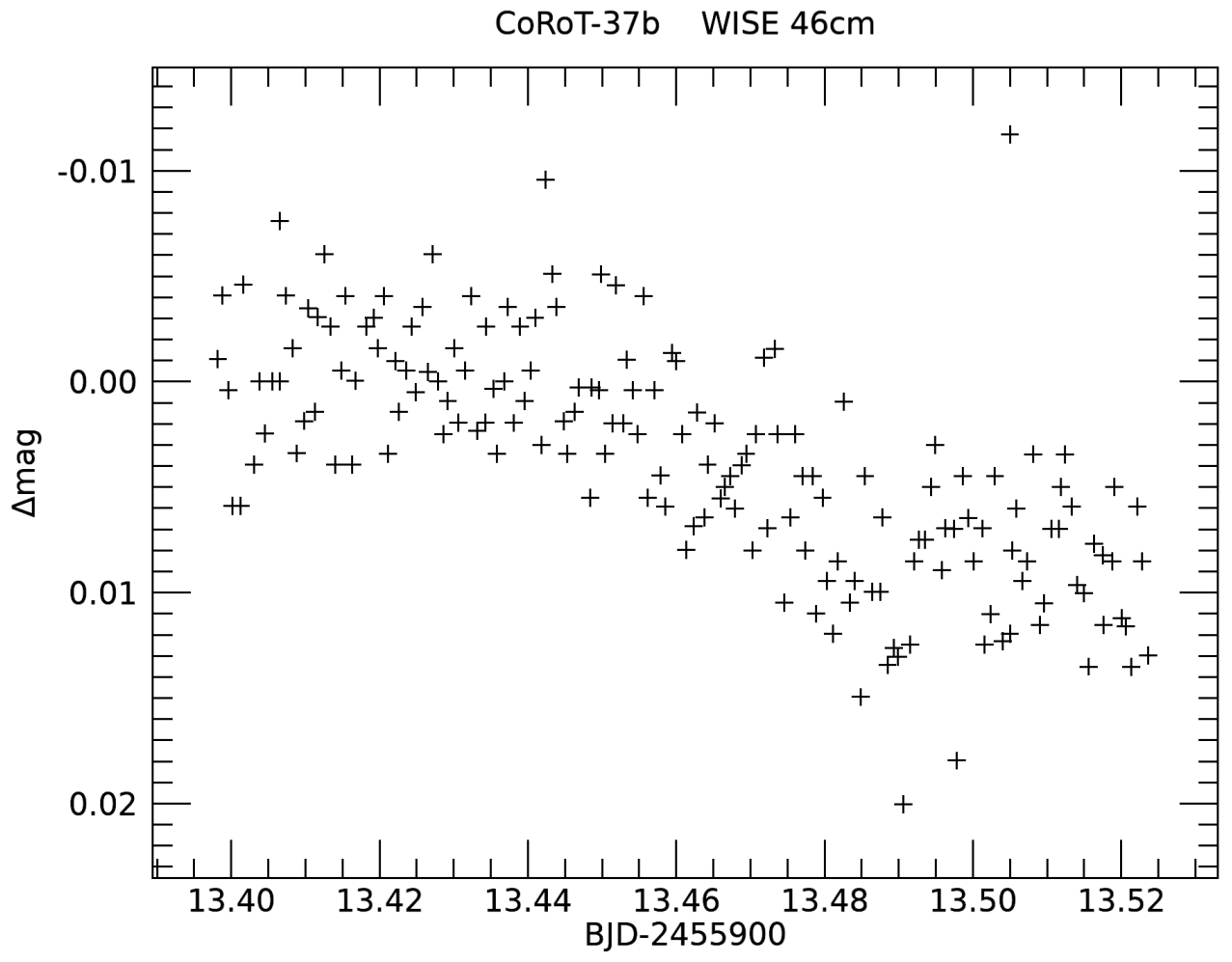

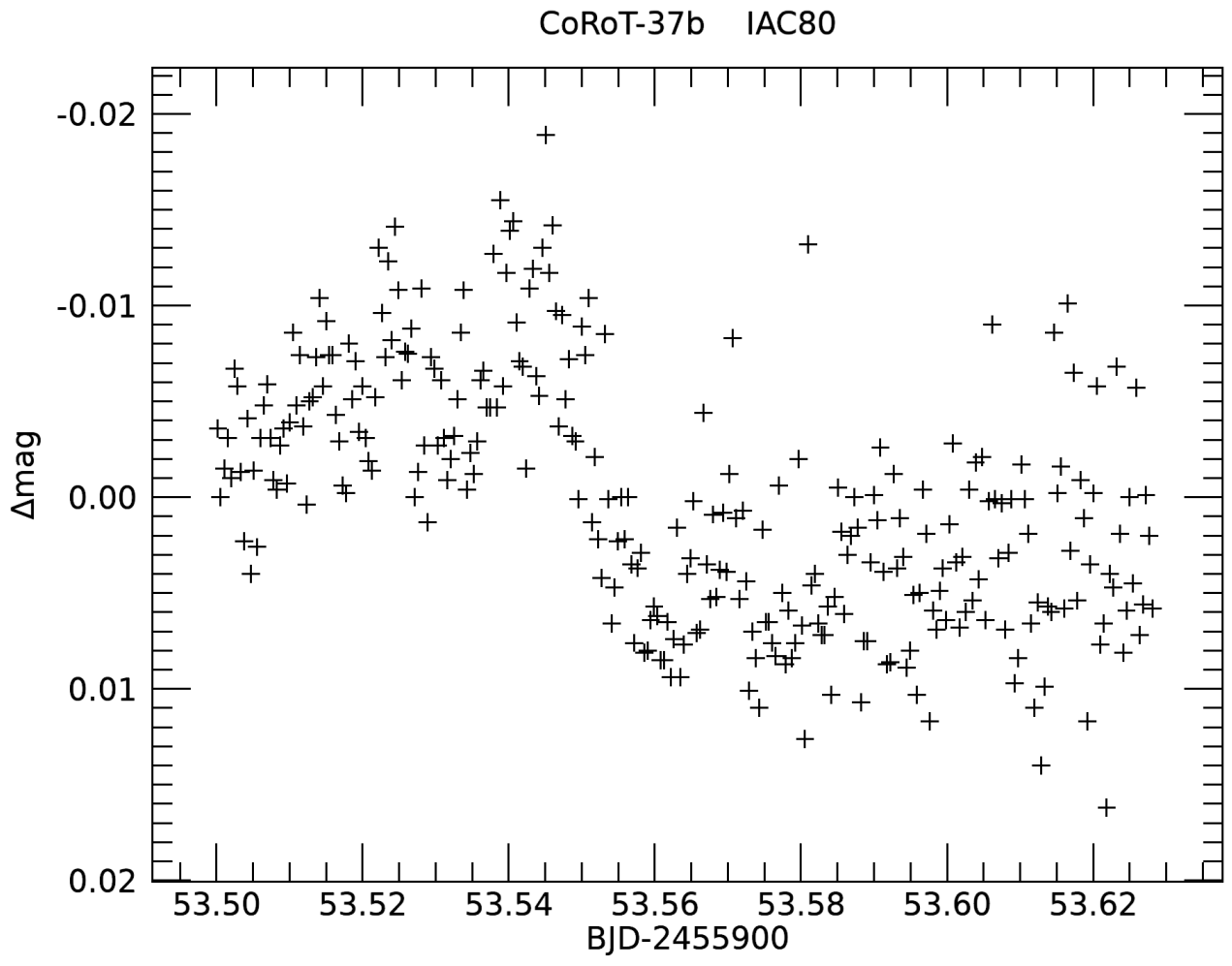

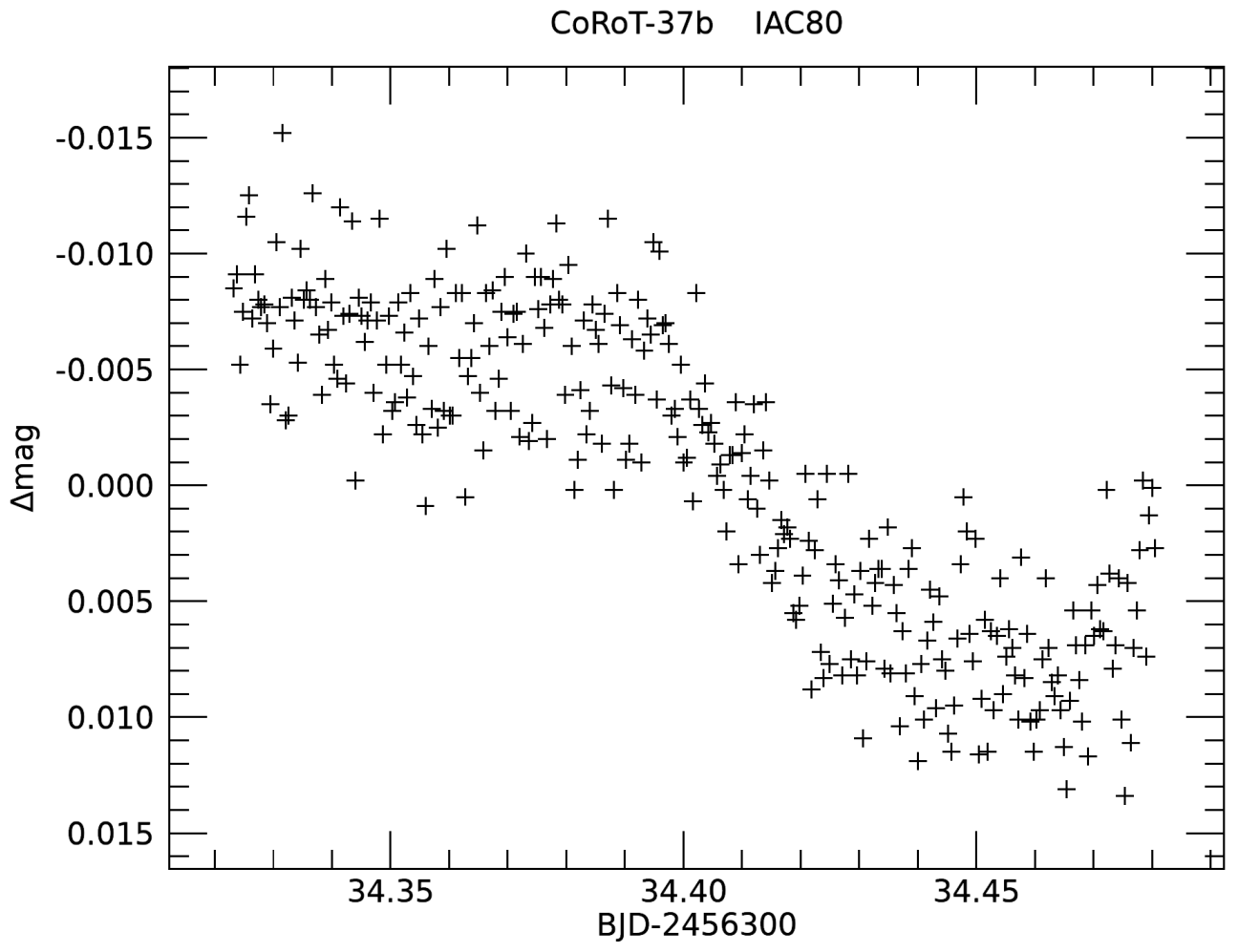